\documentclass[%
 reprint,
nofootinbib,
 amsmath,amssymb,
 aps,
]{revtex4-1}
\usepackage{amsmath}
\usepackage{graphicx}
\usepackage{dcolumn}
\usepackage{bm}
\usepackage{tabularx}
\usepackage[colorlinks=true]{hyperref}
\usepackage{color}
\usepackage[normalem]{ulem}
\usepackage{url}
\newcommand\be{\begin{equation}}
\newcommand\ba{\begin{eqnarray}}
\newcommand\ee{\end{equation}}
\newcommand\ea{\end{eqnarray}}
\newcommand\bw{\begin{widetext}}
\newcommand\ew{\end{widetext}}
\newcommand{\tol}{{\mbox{\tiny Tol}}}

\begin{document}


\title{Analytic I-Love-C relations for realistic neutron stars}
\author{Nan Jiang}
\author{Kent Yagi}
\affiliation{Department of Physics, University of Virginia, Charlottesville, Virginia 22904, USA}

\date{\today}

\begin{abstract}
Recent observations of neutron stars with radio, X-rays and gravitational waves have begun to constrain the equation of state for nuclear matter beyond the nuclear saturation density. 
To one's surprise, there exist approximate universal relations connecting certain bulk properties of neutron stars that are insensitive to the underlying equation of state and having important applications on probing fundamental physics including nuclear and gravitational physics.
To date, analytic works on universal relations for \emph{realistic} neutron stars are lacking, which may lead to a better understanding of the origin of the universality.
Here, we focus on the universal relations between the compactness ($\mathcal C$), the moment of inertia ($I$), and the tidal deformability (related to the Love number), and derive analytic, approximate I-Love-C relations.
To achieve this, we construct slowly-rotating/tidally-deformed neutron star solutions analytically starting from an extended Tolman VII model that accurately describes non-rotating realistic neutron stars, which allows us to extract the moment of inertia and the tidal deformability on top of the compactness. We solve the field equations analytically by expanding them about the Newtonian limit and keeping up to 6th order in the stellar compactness. 
Based on these analytic solutions, we can mathematically demonstrate the $\mathcal{O}(10\%)$ equation-of-state variation in the I-C and Love-C relations and the $\mathcal{O}(1\%)$ variation in the I-Love relation that have previously been found numerically.
Our new analytic relations agree more accurately with numerical results for realistic neutron stars (especially the I-C and Love-C ones) than the analytic relations for constant density stars derived in previous work.
Based on these analytic findings, we attribute a possible origin of the universality for the I-C and Love-C relations to the fact that the energy density of realistic neutron stars can be approximated as a quadratic function, as is the case for the Tolman VII solution.

\end{abstract}

\maketitle


\section{Introduction}

Neutron stars (NSs) are unique astrophysical objects for studying fundamental physics, in particular strongly-interacting many-body systems in nuclear physics~\cite{Lattimer:2000nx}.  The energy density can reach up to several times the nuclear saturation density in their inner cores, which are difficult to access with terrestrial nuclear experiments.

The relation between the mass and radius of a NS depends strongly on the equation of state (EoS) of nuclear matter, the relation between energy density and pressure. This, in turn, means that one can probe the EoS by measuring the NS mass and radius independently~\cite{guver,steiner-lattimer-brown,Lattimer2014,Ozel:2015fia,Ozel:2016oaf,Steiner:2017vmg}. For example, the X-ray payload NICER at the International Space Station recently measured the mass and radius of a pulsar to $\sim 10\%$ accuracy~\cite{Riley_2019,Miller_2019,Bogdanov_2019,Bogdanov_2019b,Guillot_2019,Raaijmakers_2019}, which has been used to constrain the EoS and measure nuclear matter parameters~\cite{Raaijmakers_2019,Christian:2019qer,Jiang:2019rcw,Raaijmakers:2019dks,Zimmerman:2020eho,Dietrich:2020lps}. Mass measurements of heavy NSs also help to constrain the EoS~\cite{1.97NS,2.01NS,Thankful}.

There are several other important global observables of NSs besides mass and radius that are useful for probing nuclear physics. Moment of inertia $I$  is expected to be measured from future observations of the double pulsar system~\cite{Kramer_2009}. A useful constraint on the EoS can be made if $I$ can be determined up to about 10\%~\cite{Lattimer_2005}. Tidal deformability (related to tidal Love number), that is a linear response of an object's quadrupole moment to the external tidal field, is encoded in gravitational waves (GWs) from binary NS mergers~\cite{Hinderer_2008, PhysRevD.77.021502}. The recent GW event GW170817 provides a first measurement of the tidal deformability~\cite{Abbott_2018}, which has been used to constrain the EoS and nuclear matter parameters~\cite{LIGO:posterior,Annala:2017llu,Abbott:2018exr,Raithel:2018ncd,Lim:2018bkq,Bauswein:2017vtn,De:2018uhw,Most:2018hfd,Annala:2019puf,Malik2018,Zack:nuclearConstraints,Carson:2019xxz,Raithel:2019ejc}.

Unlike the mass-radius relation which depends strongly on the underlying EoS, there exist universal relations among certain NS observables that do not depend sensitively on the stellar internal structure (see e.g.~\cite{YAGI20171,Doneva:2017jop} for recent reviews). One example is among
 the moment of inertia $I$, the tidal Love number 
(or the tidal deformability), and the quadrupole moment Q, which are commonly known as the I-Love-Q relations~\cite{Yagi365, Yagi:2013awa}. The relations are universal with an EoS-variation of $\sim 1\%$. Similar relations were discovered among tidal parameters in gravitational waveforms from binary NSs, known as the binary Love relations~\cite{Yagi_2016,Yagi:2016qmr}.  These relations can be used to break degeneracies among parameters in GW signals, enhancing the measurability of the tidal effects directly related to the EoS~\cite{Abbott_2018}. The relations  have many other interesting applications, including astrophysics, gravitational physics and cosmology~\cite{Yagi365, Yagi:2013awa,Yagi_2016,Yagi:2016qmr}. Reference~\cite{Yagi:2014qua} proposed emergent, approximate symmetry related to isodensity contours inside NSs as a possible origin of the universality. Reference~\cite{Sham_2015} proposed that the universality is due to the realistic EoSs being ``close'' to the incompressible, constant density one, which agrees with the picture of~\cite{Yagi:2014qua} since the approximate symmetry becomes exact in the incompressible limit. 

Most of the works on NS universal relations focused on numerical calculations, though there are  limited number of works within analytic framework. The amount of universality in the I-Love relation was analytically explained in the Newtonian limit by comparing the $n=0$ and $n=1$ polytropes~\cite{Yagi365, Yagi:2013awa}. Reference~\cite{PhysRevD.91.044017} derived the analytic, I-Love relation for constant density stars that is applicable to both the Newtonian and relativistic regimes. The authors studied the relation between the moment of inertia and compactness (the I-C relation) and that between the tidal deformability and compactness (the Love-C relation) in terms of a series expansion in compactness. The authors then eliminated the compactness to find the analytic I-Love relation. Since constant density stars are more appropriate for describing quark stars, what is still missing is the analytic, I-Love-C relations for \emph{realistic} NSs and study analytically the amount of the EoS-variation in these relations.

One useful, analytic model of non-rotating realistic NSs is the Tolman VII solution~\cite{PhysRev.55.364,Lattimer:2000nx}. This is a two-parameter (mass and radius) solution to the Einstein equations, in which the energy density profile inside a star is approximated by a quadratic function in a radial coordinate. In~\cite{PhysRevD.99.124029}, we improved this model further by extending the energy density profile to a quartic function. We introduced a phenomenological parameter $\alpha$ to more accurately capture the realistic energy density profiles. We found that this modified Tolman VII model has an improved accuracy for describing the metric and pressure or energy density functions over the original Tolman VII solution by a factor of 2--5.

In this paper, we construct analytic, I-Love-C relations for realistic NSs by extending non-rotating solutions for the modified Tolman VII model to slowly-rotating or tidally-deformed configurations, which allows us to extract the moment of inertia and tidal deformability. We treat the rotation or tidal deformation to be a small perturbation and keep only to leading order in such a perturbation. This is a good approximation since e.g. the rotation of the primary pulsar in the double pulsar binary is much smaller than the break-up rotation. Following~\cite{PhysRevD.91.044017}, we find a series-expanded solution for the moment of inertia and the tidal deformability in terms of compactness. We then eliminate the compactness to find the I-Love relation for the modified Tolman VII model applicable to realistic NSs. Varying the phenomenological parameter $\alpha$ within a reasonable range for realistic NSs allows us to analytically estimate the amount of the universality in the I-Love-C relations.

\subsection{Executive Summary}

We first constructed analytic I-C and Love-C relations for the original Tolman VII solution in terms of a series expansion in compactness. After resumming the series via Pad\'e method as done in~\cite{Sham_2015} for constant density stars, we arrived at 
\begin{eqnarray}
 \label{Pade I first}
\bar{I}_{\mathrm{Tol}}&=& \frac{\sum_{j=0}^{3} c_j^{(\bar{I})} \mathcal{C}^j}{2 \mathcal{C}^3 \sum_{i=0}^{3} d_i^{(\bar{I})} \mathcal{C}^i}, \\
\bar{\lambda}&= & \frac{16}{15}  (1-2\mathcal{C})^2 [ 2+ 2 \mathcal{C} (y_R-1) - y_R] \nonumber\\
                                &&/ \{2 \mathcal{C} [6-3 y_R +3 \mathcal{C}(5y_R-8)]\nonumber \\
                                &&+4 \mathcal{C}^3[13-11 y_R + \mathcal{C}(3y_R-2) + 2 \mathcal{C}^2(1+y_R)]\nonumber\\
                                &&+3(1-2\mathcal{C})^2[2- y_R +2 \mathcal{C}(y_R - 1)] \mathrm{ln}(1-2\mathcal{C})\}, \nonumber \\
\label{eq:k2}
\end{eqnarray}
with
\begin{equation}
 \label{Pade Love first}
y_{R,\mathrm{Tol}} = \frac{\sum_{j=0}^{2} c_j^{(\bar{\lambda})} \mathcal{C}^j}{ \sum_{i=0}^{2} d_i^{(\bar{\lambda})} \mathcal{C}^i}, 
\end{equation}
where the coefficients $c_i$ and $d_i$ are given in Table~\ref{tab:coeff}. We also present these analytic relations in a supplemental Mathematica notebook~\cite{github}.
Here the subscript ``Tol'' stands for the original Tolman VII solution.

\renewcommand{\arraystretch}{1.5}
\begin{table} [h!]
\begin{tabular}{c c }
\hline
\hline
Coefficient & Value \\
\noalign{\smallskip}
\hline
\noalign{\smallskip}
$c_0^{(\bar{I})}$ & 0 \\
$c_1^{(\bar{I})}$ & $\frac{4}{7}$\\
$c_2^{(\bar{I})}$ & $-\frac{4475212734657724923440}{3819744665891770040271}$\\ 
$c_3^{(\bar{I})}$ & $\frac{52512054644310254173804264}{119920883785672120414308045}$\\ 
\noalign{\smallskip}
\hline
$d_0^{(\bar{I})}$ & 1\\
$d_1^{(\bar{I})}$ & $-\frac{52610171361202024028}{16535691194336666841}$\\
$d_2^{(\bar{I})}$ & $ \frac{7315710780062174885366}{2662246282288203361401}$\\
$d_3^{(\bar{I})}$& $ -\frac{24682764601232290703083888}{47294805204849932715288765}$\\
\noalign{\smallskip}
\hline
$c_0^{(\bar{\lambda})}$ &$\frac{75974923394}{756262478125}$ \\
$c_1^{(\bar{\lambda})}$ & $\frac{37520415660320803457514031332847380947962712}{4113772401871789717923024281035002577115625} $\\
$c_2^{(\bar{\lambda})}$ & $\frac{959692101525501001392591086078567807004592397}{9214850180192808968147574389518405772739000}$\\ 
\noalign{\smallskip}
\hline
$d_0^{(\bar{\lambda})}$ & 1 \\
$d_1^{(\bar{\lambda})}$ & $\frac{125977972708997462470931885904739}{1673725935459728881079142766212}$\\
$d_2^{(\bar{\lambda})}$ & $-\frac{11560237280939583134679473198513569}{70296489289308613005323996180904}$\\ 
\noalign{\smallskip}
\hline
\hline
\end{tabular}
\caption{\label{tab:coeff}
Coefficients $c^{(\bar{I})}_i$, $d^{(\bar{I})}_i$ in Eq.~\eqref{Pade I first} and $c^{(\bar{\lambda})}_i$, $d^{(\bar{\lambda})}_i$ in Eq.~\eqref{Pade Love first}. }
\end{table}

\begin{figure}[htp]
\includegraphics[width=8.5cm]{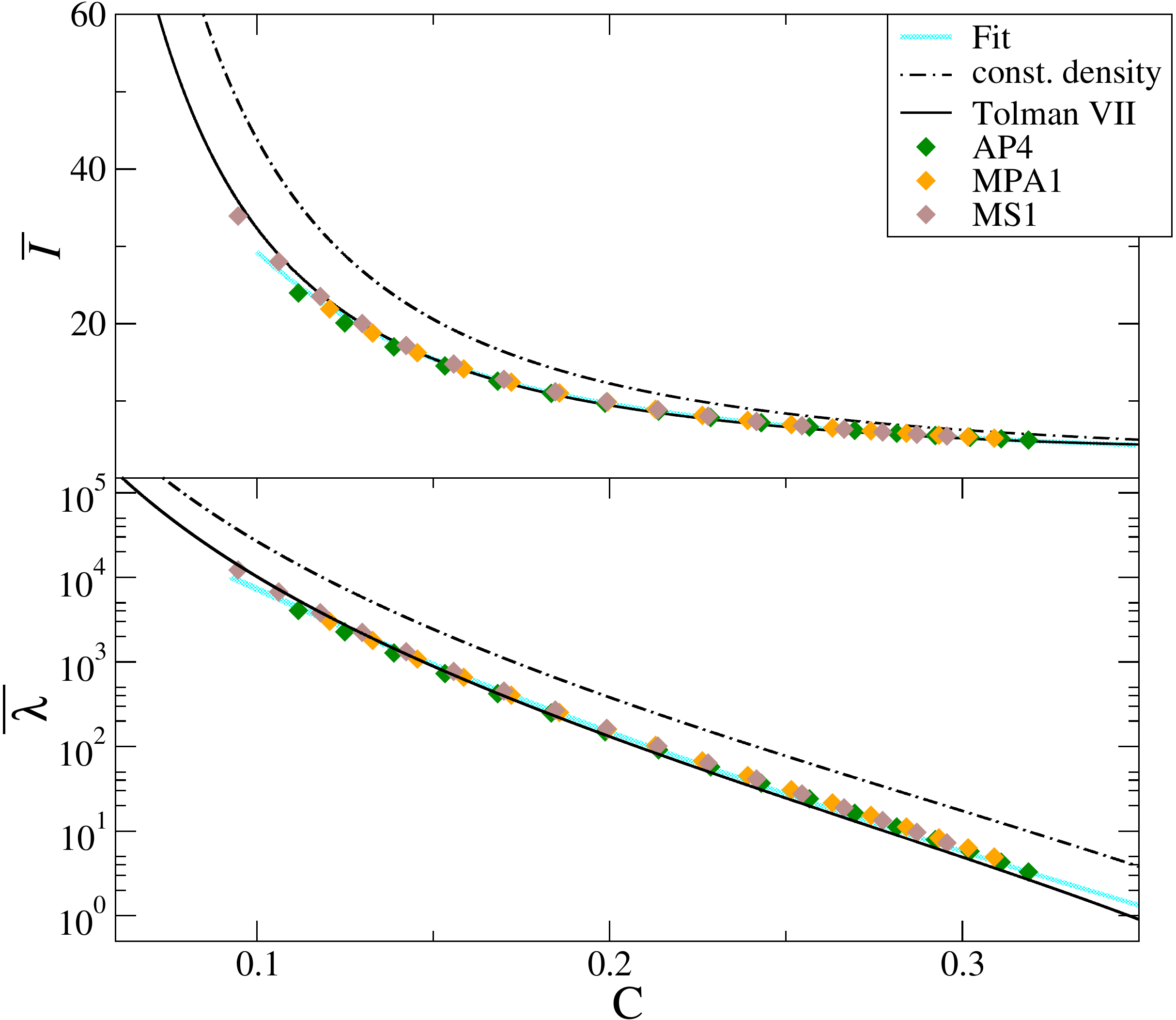}
\caption{Approximate, analytic I-C relation (top) and Love-C relation (bottom) for the original Tolman VII model (solid) for realistic NSs and constant-density-star model~\cite{PhysRevD.91.044017} (dotted-dashed), both in  Pad\'e resummed form.  We also present those for realistic EoSs representing soft (AP4), intermediate (MPA1) and stiff (MS1) EoS classes.   We can see a clear improvement in accuracy for the Tolman-VII curves over the constant density ones in both panels. For comparison, the numerical fit curves in~\cite{YAGI20171} are plotted in cyan.
}
\label{fig:PadeTolman}
\end{figure}

Figure~\ref{fig:PadeTolman} compares these analytic relations with numerical ones for NSs with realistic EoSs and the analytic one for constant density stars. Observe that the analytic relations for the Tolman VII model can accurately describe the realistic relations. 
We can see the analytic curves are very close to the numerical fit ones in~\cite{YAGI20171}, with fractional differences $\sim$ 1\% for the I-C curve and several percent for the Love-C curve.
Observe also that the former is much more accurate than the one for constant density stars which is more appropriate for quark stars. We next constructed the I-Love relation semi-analytically for the original Tolman VII solution by eliminating the compactness from the I-C and Love-C relations, which again accurately models the numerical results.

We also derived the analytic I-C/Love-C and semi-analytic I-Love relations for the modified Tolman VII model~\cite{PhysRevD.99.124029}. By varying the density profile parameter $\alpha$ that corresponds to varying the EoS, we successfully obtained the 10\% EoS-variation in the I-C and Love-C relations while 1\% EoS-variation 
in the I-Love relations, which agree with previous numerical findings~\cite{Yagi365,Yagi:2013awa,YAGI20171,Maselli:2013mva,Urbanec:2013fs}.

\subsubsection{Organization}

The organization of the rest of the paper is as follows. In Sec.~\ref{sec:level1}, we briefly introduce what original and modified Tolman VII solutions are. In Secs.~\ref{sec:2} and \ref{sec:3}, we describe how we construct the analytic I-C and Love-C relations respectively, and compare them against  numerical relations for realistic NSs and analytic relations for constant density stars. We also analytically estimate the amount of universality in these relations. In Sec.~\ref{sec:4}, we present the results of the semi-analytic I-Love relation. In Sec.~\ref{sec:origin}, we use our analytic findings to present a possible origin of the universality for the I-C and Love-C relations.  In Sec.~\ref{sec:Conclusion}, we conclude  and give possible  directions for future work.  We use the geometric units of $c$ =  1 and $G$ = 1 throughout this paper.

\section{\label{sec:level1}Modified Tolman VII solution}

Here, we briefly review the original Tolman VII~\cite{PhysRev.55.364} and modified Tolman VII~\cite{PhysRevD.99.124029} solution of a static, isolated and spherically symmetric NS in the interior region. We use the metric ansatz given  by 
\begin{eqnarray}
\label{eq:metric}
ds^2 = ds_0^2 \equiv -e^\nu dt^2+ e^\lambda dr^2+r^2(d\theta^2+\sin^2\theta d\phi^2), 
\end{eqnarray}
where $\nu$ and $\lambda$ are functions of $r$.
We also assume that matter inside a NS is modeled by a perfect fluid with the stress-energy tensor given by
\begin{eqnarray}
\label{eq:matter}
T_{\mu\nu}=(\rho+p) u_\mu u_\nu+ p g_{\mu\nu}.
\end{eqnarray}
Here $\rho$ represents energy density, $p$ represents pressure and $u_{\mu}$ is the 4-velocity of the fluid.
Substituting Eqs.~\eqref{eq:metric} and~\eqref{eq:matter} into the Einstein equations, one 
finds the following three differential equations with four unknown quantities ($\nu$, $\lambda$, $p$ and $\rho$, which are functions of $r$ only)  as~\cite{PhysRev.55.364}
\begin{equation}
\label{eq:diff-eq-nu}
\frac{d}{dr} \left(\frac{e^{-\lambda}-1}{r^2}\right)+\frac{d}{dr}\left( \frac{e^{-\lambda} \nu'}{2 r}\right)+e^{-\lambda-\nu} \frac{d}{dr}\left (\frac{e^\nu \nu'}{2 r}\right) = 0,
\end{equation}
\begin{eqnarray}
e^{-\lambda} \left(\frac{\nu'}{r}+\frac{1}{r^2}\right)-\frac{1}{r^2} =8 \pi p,
\label{pressure formula}
\end{eqnarray}
\begin{eqnarray}
\label{eq:m}
\frac{dm}{dr} = 4\pi r^2 \rho, 
\end{eqnarray}
where a prime denotes a derivative with respect to $r$ and
\begin{equation}
\label{eq:e_lambda}
e^{-\lambda} \equiv 1 - \frac{2m(r)}{r}
\end{equation}
with $m(r)$ representing the mass enclosed in a sphere of radius $r$.

 To make the system of equations 
 closed, one typically specifies an EoS relating $p$ as a function of $\rho$. Instead, in this paper, we specify the radial profile of $\rho$, which corresponds to choosing an EoS. 

In the exterior region $(r>R)$, $p_\mathrm{ext} = 0$, $\rho_\mathrm{ext} = 0$ and $m(r) = M$ with the stellar mass $M$. Then, we have
\begin{equation}
e^{\nu_\mathrm{ext}} =e^{-\lambda_\mathrm{ext}} =1 - \frac{2M}{r}\,.
\end{equation}

\subsection{\label{sec:Tolman}Original Tolman VII solution}

In the original Tolman VII solution~\cite{PhysRev.55.364},  
the energy density profile is given by
\begin{eqnarray}
\label{eq:rho-Tol}
\rho_\tol(\xi) = \rho_{c} (1-\xi^2).
\end{eqnarray}
Here the dimensionless radial coordinate $\xi$ is given by $\xi=r/R$ with the stellar radius $R$ 
and $\rho_c$ is the central energy density. The subscript ``Tol''  indicates the quantity for the original Tolman VII solution. We next substitute the above density profile  into Eq.~\eqref{eq:m} and integrate over $r$ under the boundary condition $m(0)=0$. One finds
\begin{eqnarray}
\label{eq:m-Tol}
m_\tol(r)  =  4 \pi \rho_{c} \left(\frac{r^3}{3} - \frac{r^5}{5 R^2}\right).
\end{eqnarray}
$\rho_c$ can be expressed in terms of  $R$ and the stellar compactness $\mathcal{C} \equiv M/R $ as
\begin{equation}
\rho_c = \frac{15 \mathcal{C}}{8 \pi  R^2}.
\end{equation} 
We can substitute this into Eqs.~\eqref{eq:rho-Tol} and~\eqref{eq:m-Tol} to find
\begin{eqnarray}
\label{eq:original Tol}
\rho_\tol(\xi) & =& \frac{15 \mathcal{C}}{8 \pi R^2}  (1- \xi^2), \\
\label{original m}
m_\tol(\xi) &=& \mathcal{C} R \left(\frac{5}{2} \xi^3 - \frac{3}{2}\xi^5\right).
\end{eqnarray}
Using Eq.~\eqref{eq:e_lambda}, $e^{-\lambda}$ is given by
\begin{eqnarray}
e^{-\lambda_\tol(\xi)} &=& 1- \mathcal{C} \xi^2 \left(5 - 3 \xi^2\right).
\label{exp(-l) expression}
\end{eqnarray}
Next, one can solve 
for $\nu$ and $p$ to find
\begin{eqnarray}
e^{\nu_\tol(\xi)} &=& C_1^\tol \cos^2\phi_\tol, \\
\label{original nu}
p_\tol (\xi)&=& \frac{1}{4 \pi R^2} \left[ \sqrt{3 \mathcal{C} e^{- \lambda_\tol}} \tan\phi_\tol- \frac{\mathcal{C}}{2} (5 - 3 \xi^2)\right], \nonumber \\
\label{original p}
\end{eqnarray}
with 
\begin{eqnarray}
\phi_\tol (\xi) &=& C_2^\tol - \frac{1}{2} \log\left(\xi^2 - \frac{5}{6} +\sqrt{\frac{e^{-\lambda_\tol}}{3 \mathcal{C}}}\right),
\label{eq:phi_orig}
\end{eqnarray}
and the integration constants are given by
\begin{eqnarray}
C_1^\tol &=& 1 - \frac{5 \mathcal{C}}{3}, \\
C_2^\tol &=& \arctan \sqrt{\frac{\mathcal{C}}{3 (1-2 \mathcal{C})}} + \frac{1}{2}\log\left(\frac{1}{6} + \sqrt{\frac{1 - 2\mathcal{C}}{3 \mathcal{C}}}\right). \nonumber \\
\end{eqnarray}
The above solution is the so-called Tolman VII solution.

\subsection{\label{sec:modTolman}Modified Tolman VII solution}

\begin{figure}[htp]
\includegraphics[width=8.5cm]{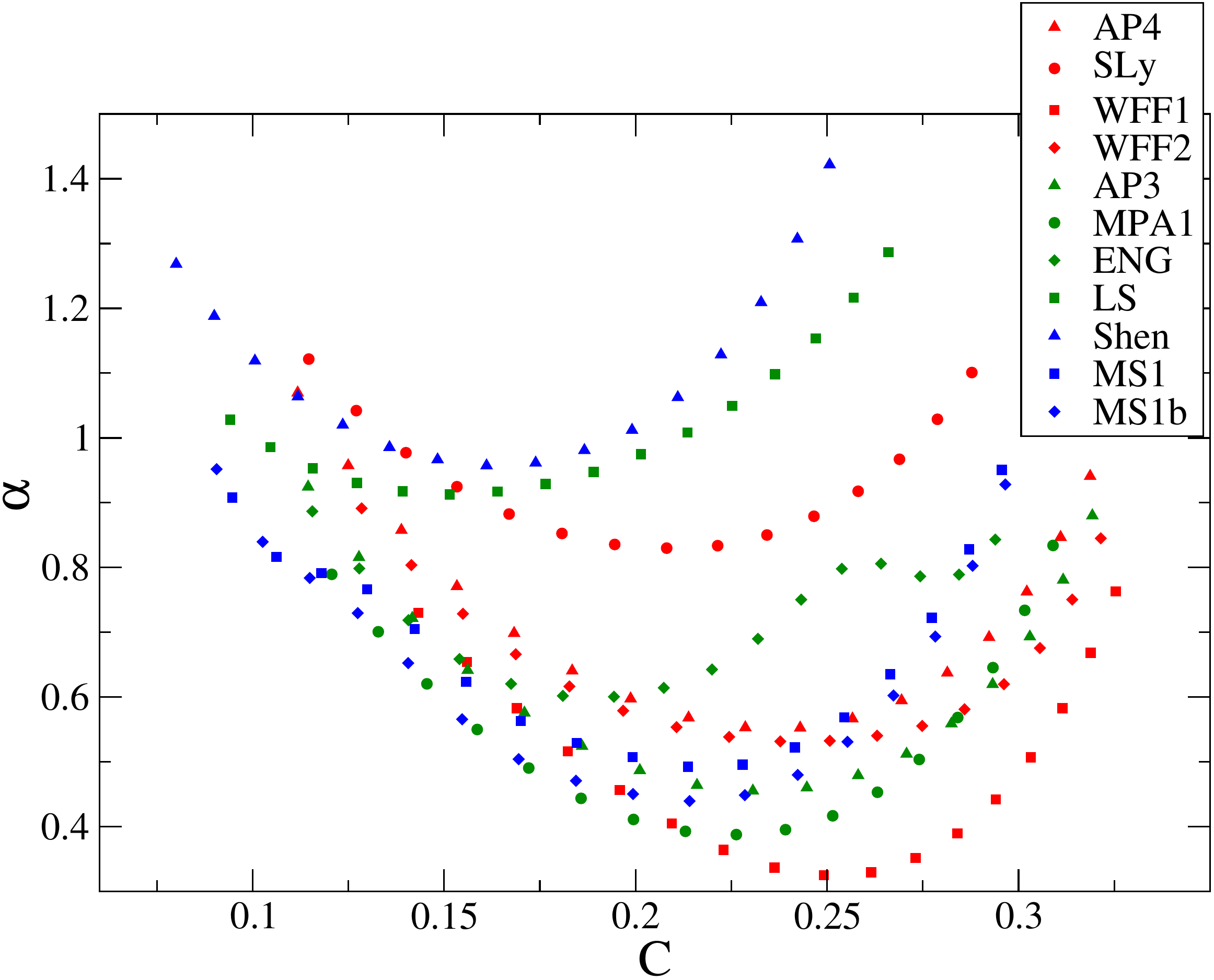}
\caption{The parameter $\alpha$ introduced in the density profile of the modified TolmanVII model (Eq.~\eqref{eq:rho_imp}) against the compactness $\mathcal{C}$
for 11 realistic EoSs. The red, green and blue dots represent soft, intermediate and stiff EoSs respectively.
}
\label{fig:alpha-C}
\end{figure}

Modified Tolman VII model~\cite{PhysRevD.99.124029} introduces an additional term to Eq.~\eqref{eq:rho-Tol} with a free parameter $\alpha$ in order to capture the variation in the energy density profile among different EoSs:
\begin{equation}
\label{eq:rho_imp}
\rho_\mathrm{mod}(\xi) = \rho_{c} \left[1- \alpha \xi^2 + (\alpha - 1) \xi^4\right].
\end{equation}
Obviously, this new profile reduces to the original one 
in the limit $\alpha \to 1$.  
Figure~\ref{fig:alpha-C} presents $\alpha$ obtained by fitting a numerical energy density profile with Eq.~\eqref{eq:rho_imp} for various EoSs\footnote{The 11 EoSs considered here are same as those in~\cite{PhysRevD.99.124029}, which all support NSs with their mass above $2M_\odot$.} and compactness. Observe that $\alpha$ ranges in $\alpha \in $[0.4,1.4] when $\mathcal{C} \in$[0.05,0.35]. We will vary $\alpha$ within this range in later calculations.

The remaining equations are modified from the original Tolman VII one as follows.
$m$ and $\lambda$ now become
\begin{eqnarray}
m_\mathrm{mod}(\xi) &=& 4 \pi \rho_c R^3 \xi^3 \left(\frac{1}{3} - \frac{\alpha}{5} \xi^2 +\frac{\alpha - 1}{7} \xi^4\right), \\
e^{-\lambda_\mathrm{mod}(\xi)} &=& 1- 8 \pi R^2 \xi^2 \rho_c \left(\frac{1}{3} -\frac{\alpha}{5} \xi^2+ \frac{\alpha-1}{7} \xi^4 \right),
\label{mod m}
\end{eqnarray}
with the subscript ``mod'' representing the quantities in the modified Tolman solution.
The expressions for $\nu$ and $p$ are then adjusted to be 
\begin{widetext}
\begin{eqnarray}
e^{\nu_\mathrm{mod}(\xi)} &=& C_1^\mathrm{mod} \cos^2\phi_\mathrm{mod}, \\
p_\mathrm{mod}(\xi) &=& \sqrt{\frac{e^{-\lambda_\tol} \text{$\rho $}_c}{10
   \pi }
   } \frac{\tan \phi_\mathrm{mod}}{R}
+\frac{1}{15} \left(3 \text{$\xi
   $}^2-5\right) \text{$\rho $}_c +\frac{6
   (1-\alpha ) \text{$\rho $}_c}{16 \pi  (10-3
   \alpha) \text{$\rho $}_c
   R^2-105},
   \end{eqnarray}
with
\begin{eqnarray}
\phi_\mathrm{mod} (\xi)&=& C_2^\mathrm{mod} - \frac{1}{2} \log\left(\xi^2 - \frac{5}{6} +\sqrt{\frac{5 e^{-\lambda_\tol}}{8 \pi R^2 \rho_c}}\right), \\
C_1^\mathrm{mod}  &=& (1-2\mathcal{C})  \left\{ 1+\frac{8 \pi R^2 \rho_c (10 -3\alpha)^2 (15 - 16 \pi R^2 \rho_c)}{3 [105+16\pi R^2  \rho_c (3\alpha-10)]^2}\right\}, 
   \\
C_2^\mathrm{mod} &=& \arctan \left[-\frac{2 (10-3 \alpha )
  R \sqrt{6 \pi  \text{$\rho $}_c  \left(15-16 \pi 
   \text{$\rho $}_c R^2\right)}}{48 \pi
    (10-3 \alpha) \text{$\rho $}_c
   R^2-315}\right]  + \frac{1}{2}\log\left(\frac{1}{6} + \sqrt{\frac{5}{8 \pi  \text{$\rho $}_c
   R^2}-\frac{2}{3}}\right).
\end{eqnarray}

\end{widetext}

Notice that the above solution is parameterized by ($\mathcal{C}, R, \rho_c$ and $\alpha$). We can further eliminate $\rho_c$ from the condition $M = m_\mathrm{mod}(1)$, which yields
\begin{eqnarray}
\rho_c = \frac{105  \mathcal{C}}{8 \pi R^2 (10 - 3 \alpha)}.
\label{eq:rhoc}
\end{eqnarray}
Using this into the modified Tolman VII expressions, we obtain a three-parameter ($\mathcal{C}, R, \alpha$) (or equivalently $(M,R,\alpha)$) model which will be used in the following calculations. 

\section{\label{sec:2}Moment of Inertia}

\subsection{Formulation}

One can extract the moment of inertia $I$ from the asymptotic behavior of the time-spatial component of the metric at infinity for a slowly rotating NS. For simplicity, we consider uniform rotation. The metric ansatz is given by
\begin{equation}
ds^2 = ds_0^2 -2\Omega(1-\omega)r^2 \sin^2\theta dt d\phi,
\end{equation}
where $ds_0^2$ is the non-rotating part of the line element in Eq.~\eqref{eq:metric}, $\Omega$ is the constant stellar angular velocity and $\omega$ is a function of $r$ (or $\xi$). Substituting this ansatz into the Einstein equations, one finds an equation for $\omega$ as~\cite{1967ApJ150.1005H}
\begin{eqnarray}
\frac{d}{d\xi} \left(\xi^4 j \frac{d \omega}{d\xi} \right) + 4 \xi^3 \frac{dj}{d\xi} \omega = 0,
\label{eq:I}
\end{eqnarray}
with the boundary condition
\begin{eqnarray}
\omega  \mid _{\xi \to 0} = 1, \quad \frac{d\omega}{d \xi} \bigg| _{\xi \to 0}= 0,
\end{eqnarray}
and $j(\xi)= e^{(-\lambda+\nu)/2}$.  

Let us next look at the exterior solution. Integrating the equation in the exterior region (with $j(\xi)=1$), one finds
\begin{eqnarray}
\omega_\mathrm{ext}(r) = 1- \frac{2 I }{r^3}.
\end{eqnarray}
The closed form of $I$ can be obtained easily from the differential equations and its boundary conditions above, which is~\cite{1967ApJ150.1005H}
 
\begin{eqnarray}
\label{eq:I_integ}
I = \frac{8 \pi R^5}{3} \int_{0}^{1} \frac{ \xi^5  ( p+\rho) e^{-(\nu+\lambda)/2} \omega }{\xi-\frac{2 m}{R}} d\xi.
\end{eqnarray}

\subsection{\label{sec:modTolman ana I}Analytic Solutions}

 We now find an analytic solution to Eq.~\eqref{eq:I} for the modified Tolman VII solution. Unfortunately, we were not able to find an exact solution. Instead, we apply a post-Minkowskian recursive perturbation method adopted in Chan \textit{et al}.~\cite{PhysRevD.91.044017} for incompressible stars to derive an approximate, analytic solution for $\omega$ and $I$.
 
 We begin by defining
\begin{equation}
\tilde \omega(\xi) \equiv 1 - \omega(\xi),
\end{equation}
 and expanding $\tilde \omega(\xi)$ and $j(\xi)$ in power series of $\mathcal{C}$: 
\begin{eqnarray}
\label{eq:expI}
\tilde \omega(\xi,\mathcal{C}) \approx \sum_{n=0}^N \tilde \omega_n(\xi) \mathcal{C}^n, \quad
j(\xi,\mathcal{C}) \approx \sum_{n=0}^N j_n(\xi) \mathcal{C}^n.
\end{eqnarray}
Here $ \tilde \omega_n(\xi)$ and $j_n(\xi)$ 
are functions of $\xi$ only while $N$ corresponds to the order of the expansion that we keep. In the Newtonian limit ($\mathcal{C} \to 0$), 
one finds
\begin{eqnarray}
j_0(\xi) = 1, \quad \tilde \omega_0(\xi) = 0.
\end{eqnarray}

\begin{figure}[htp]
\includegraphics[width=8.5cm]{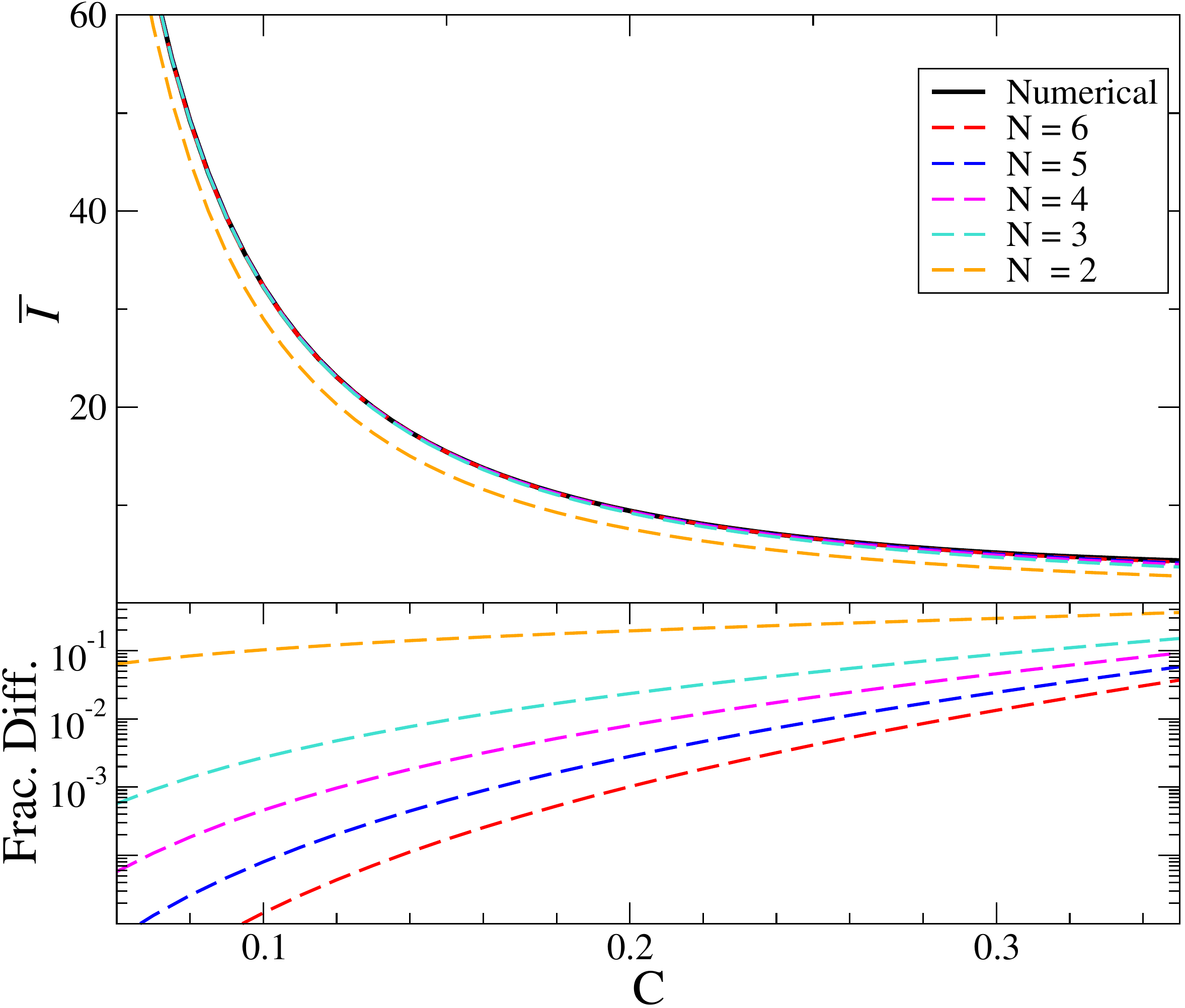}
\caption{(Top) I-C relation for the original TolmanVII solution with different orders of series expansion in $\mathcal{C}$. (Bottom) Fractional differences between the curve obtained numerically and the analytic one at each expansion order. Observe that the analytic relation converges as we increase the order.
}
\label{fig:IC converge}
\end{figure}

Having the above series expansion at hand, we can solve the differential equation. Namely, we substitute Eq.~\eqref{eq:expI} into Eq.~\eqref{eq:I} and solve the latter for $\tilde \omega_n$ order by order in powers of $\mathcal{C}$. We can then derive the series expansion of the dimensionless moment of inertia from Eq.~\eqref{eq:I_integ} as
\begin{equation}
\bar{I} \equiv \frac{I}{M^3} \approx \sum_{n=0}^N \bar I_n \mathcal{C}^{n-2}.
\end{equation}
The series starts with $\mathcal{C}^{-2}$ since $I \propto M R^2$ in the  Newtonian limit. $\bar I$ is given in terms of $\tilde \omega$ as 
\begin{equation}
\bar I = \frac{\tilde \omega_R}{2\mathcal{C}^3}, \quad \tilde \omega_R \equiv \tilde \omega(\xi=1).
\end{equation}
Figure~\ref{fig:IC converge} shows how the I-C series converges as one increases the order of expansion. Observe that the expansion to 6th order in compactness can accurately describe the numerical relation.
Since the expressions for $\bar{I}_n$ are lengthy and not illuminating, we show them in a supplemental Mathematica notebook~\cite{github}.

\begin{figure}[htp]
\includegraphics[width=8.5cm]{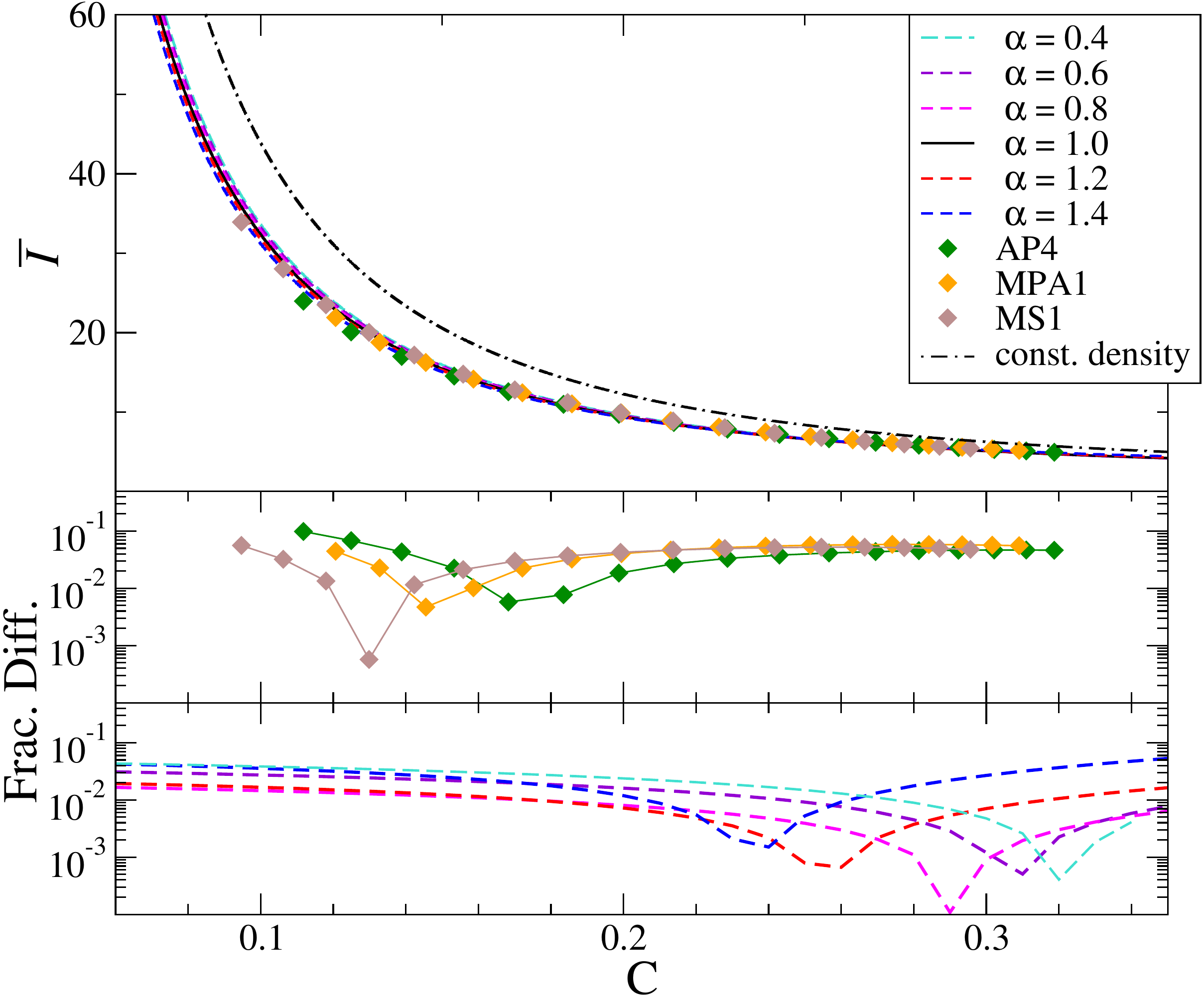}
\caption{(Top) Approximate, analytic I-C relation (up to sixth order in the series expansion in $\mathcal{C}$) for the modified Tolman VII solution with five representative values of $\alpha$. For comparison, we also show the numerical relation for three representative realistic EoSs, namely AP4 (soft), MPA1 (intermediate) and MS1 (stiff). The  dotted-dashed line represents the relation for incompressible constant density stars in~\cite{PhysRevD.91.044017}. (Middle) Relative fractional difference of the I-C relation for the realistic EoSs from that for the original Tolman solution. (Bottom) Relative fractional difference for the modified Tolman VII relation from the original one. Observe that the EoS-variation  is kept under 10\% for both realistic EoSs and modified Tolman VII relations, in agreement with that for realistic EoSs in Fig.~15 of~\cite{YAGI20171}. 
}
\label{fig:I-C}
\end{figure}

We present in the top panel of Fig.~\ref{fig:I-C} the approximate, analytic I-C relation for the modified Tolman VII solution. For comparison, we also show the approximate, analytic relation for incompressible stars~\cite{PhysRevD.91.044017} and the relations for some representative realistic EoSs. Observe that the new analytic relation can beautifully approximate the numerical relation for the realistic EoSs.  Observe also that the relation for incompressible stars found in~\cite{PhysRevD.91.044017} does not provide an accurate description for realistic NSs\footnote{The relation for incompressible stars can accurately describe the relation for quark stars~\cite{YAGI20171}.}. The new relation found in this paper gives us the first analytic expression for the I-C relation for realistic NSs other than the fits in e.g.~\cite{Breu:2016ufb,Staykov:2016mbt,YAGI20171}.

The bottom panel shows the fractional difference of the modified Tolman VII relation from the original one. Notice that when we vary $\alpha$ within the range for realistic EoSs, the analytic relation varies by $\sim 10\%$. This amount agrees with that of the EoS-variation in the relation for realistic NSs in the middle panel (see also~\cite{YAGI20171}). Thus, the new relation provides us an analytic explanation for the amount of universality in the I-C relation.

Let us investigate further the EoS-variation in the I-C relation using the new analytic I-C relation. Expanding $\bar I_n$ about $\alpha=1$ (the original Tolman solution), we find
\begin{align*}
  \bar I_0 &= 0.286 [ 1 - 0.105 (\alpha - 1) + \mathcal{O}(\alpha - 1)^2 ]\\
  \bar I_1 &= 0.323 [ 1 - 0.100 (\alpha - 1) + \mathcal{O}(\alpha - 1)^2 ]\\
  \bar I_2 &= 0.462 [ 1 + 0.258(\alpha - 1) + \mathcal{O}(\alpha - 1)^2 ]\\
  \bar I_3 &= 0.732 [ 1 + 0.403(\alpha - 1) + \mathcal{O}(\alpha - 1)^2 ]\\
  \bar I_4 &= 1.226 [ 1 + 0.545(\alpha - 1) + \mathcal{O}(\alpha - 1)^2 ]\\
 \bar I_5 &= 2.132 [ 1 + 0.688(\alpha - 1) + \mathcal{O}(\alpha - 1)^2 ].
 \label{eq: alpha dependence I}
\end{align*}
This shows that the $\alpha$ dependence on the leading contribution to $\bar I$ ($\bar I_0$) is $\sim 10\%$ (because the relative coefficient is 0.105), and similar for $\bar I_1$. This analytically explains the origin of the $\mathcal{O}(10\%)$ EoS-variation in the relation. On the other hand, the $\alpha$ dependence becomes larger as we increase $n$ in $\bar I_n$, though such contributions are higher order and does not affect the relation much.

\section{\label{sec:3}Tidal Love Number}

\subsection{Formulation}

In this section, we present the calculation of the tidal deformability or the tidal Love number due to tidal deformation. For example, a primary NS in a binary acquires tidal deformation due to the tidal field created by a companion star. Such an effect is characterized by the dimensionless tidal deformability, which is defined as
\begin{eqnarray}
\bar{\lambda}\equiv \frac{\lambda}{M^5}\equiv -\frac{Q}{M^5} \mathcal{E} = \frac{2}{3} k_2 \mathcal{C}^{-5}.
\label{lambda}
\end{eqnarray}
Here $Q$ and $\mathcal{E}$ are the tidally-induced quadruple moment and the external tidal field 
and $ k_2 $ is the tidal Love number defined by $k_2\equiv (3/2) (\lambda/R^5)$. To calculate $Q$ and $\mathcal{E}$, we follow the formulations and conventions established in ~\cite{Hinderer_2008}. We begin by the metric ansatz
\begin{eqnarray}
ds^2 &=& ds_0^2 - [h_2 (e^\nu dt^2 + e^\lambda  dr^2) \nonumber \\
&&- r^2 K_2 (d\theta^2 + \sin^2\theta \ d\phi^2) ]Y_{2m}(\theta,\phi),
\end{eqnarray}
where $h_2(r)$ and $K_2(r)$ are quadrupolar tidal perturbations while $Y_{\ell m}$ are spherical harmonics. Substituting this into the Einstein equations, one can derive a second order differential equation for $h_2$ as~\cite{Hinderer_2008} 
\begin{eqnarray}
&&  \xi y' + y^2 + y e^\lambda [1 + 4 \pi \xi^2 R^2 (p - \rho)] \nonumber \\
      && - \left\{\frac{6 e^\lambda}{\xi^2} - 4 \pi e^\lambda R^2 \left[ 5 \rho+ 9 p + (\rho+ p) \frac{d \rho}{d p}\right] + \nu'{}^2 \right\} \xi^2 =0, \nonumber \\
\label{eq:y}
\end{eqnarray}
where 
\begin{equation}
y \equiv \frac{\xi}{h_2} \frac{d h_2}{d \xi},
\end{equation}
and the prime represents a derivative with respect to $\xi$. The initial condition is given by $y(\xi=0)=2$.
Solving the above equation, one can find $\bar{\lambda}$ using Eq.~\eqref{lambda} as in Eq.~\eqref{eq:k2} with $y_R \equiv y(\xi = 1)$~\cite{Hinderer_2008,Damour:2009vw}.

\subsection{\label{sec:modTolman ana Love}Analytic Solution}

We now study analytic expressions for $h_2$ (or $y$) and the tidal deformability. Similar to the case of the moment of inertia, we were not able to solve Eq.~\eqref{eq:y} exactly for the modified Tolman VII model. Thus, we follow the same procedure as before and consider series expansion in $\mathcal C$.

We begin by expanding $y$ as~\cite{PhysRevD.91.044017}
\begin{equation}
\label{eq:y_exp}
y(\xi, \mathcal{C}) \approx \sum_{n=0}^N  y_n (\xi) \mathcal{C}^n,
\end{equation}
where each coefficient $y_n$ is a function of $\xi$ only. Substituting this expansion into Eq.~\eqref{eq:y} and look at order by order, one can derive differential equations for $y_n$. Unlike the case of the moment of inertia, we were not able to find analytic exact solutions to these equations for the modified Tolman VII model\footnote{For the original Tolman model, $y_0$ can be solved exactly in terms of hypergeometric functions, though the differential equation is too complicated to be solved for $y_1$ and higher.}. Thus, we further expand $y_n$ about $\xi = 0$ as 
\begin{equation}
\label{eq:y_n_exp}
y_n(\xi) \approx \sum_{k=1}^K y_n^{(2k)} \xi^{2k},
\end{equation}
where $y_n^{(2k)}$ is now a constant while $K$ is the expansion order in terms of $\xi$. We expand Eq.~\eqref{eq:y} in powers of both $\mathcal C$ and $\xi$, find algebraic equations for $y_n^{(2k)}$ order by order and solve them. We show $y_n(\xi)$ expanded up to 12th order in $\xi$ ($K$ = 6) in the supplemental Mathematica notebook~\cite{github}.

Figure~\ref{fig:Love converge} compares the analytic relations between $\bar \lambda$ and $\mathcal C$ (the Love-C relations) for the original Tolman solution at different expansion order in $\mathcal{C}$ against the numerical relation. For the former, we fixed $K=6$ for the $\xi$ expansion. 
Unlike the I-C case in Fig.~\ref{fig:IC converge}, the series does not converge. We found that the sound speed squared ($dp/d\rho)$ in Eq.~\eqref{eq:y} becomes inaccurate as one goes to higher order in $\mathcal C$. The analytic relation is given in series expansion in $\mathcal C$, thus the approximation becomes worse in large  $\mathcal C$ region.
In this section, we keep up to 3rd order in $\mathcal C$ ({$N=3$ in Eq.~\eqref{eq:y_exp}) and 12th order in $\xi$ ($K=6$ in Eq.~\eqref{eq:y_n_exp}) that most accurately approximates the correct numerical relation. When deriving the I-Love relation, we will use the relation up to 6th order in $\mathcal{C}$ as we will explain in more detail in Sec.~\ref{sec:4}.

\begin{figure}[htp]
\includegraphics[width=8.5cm]{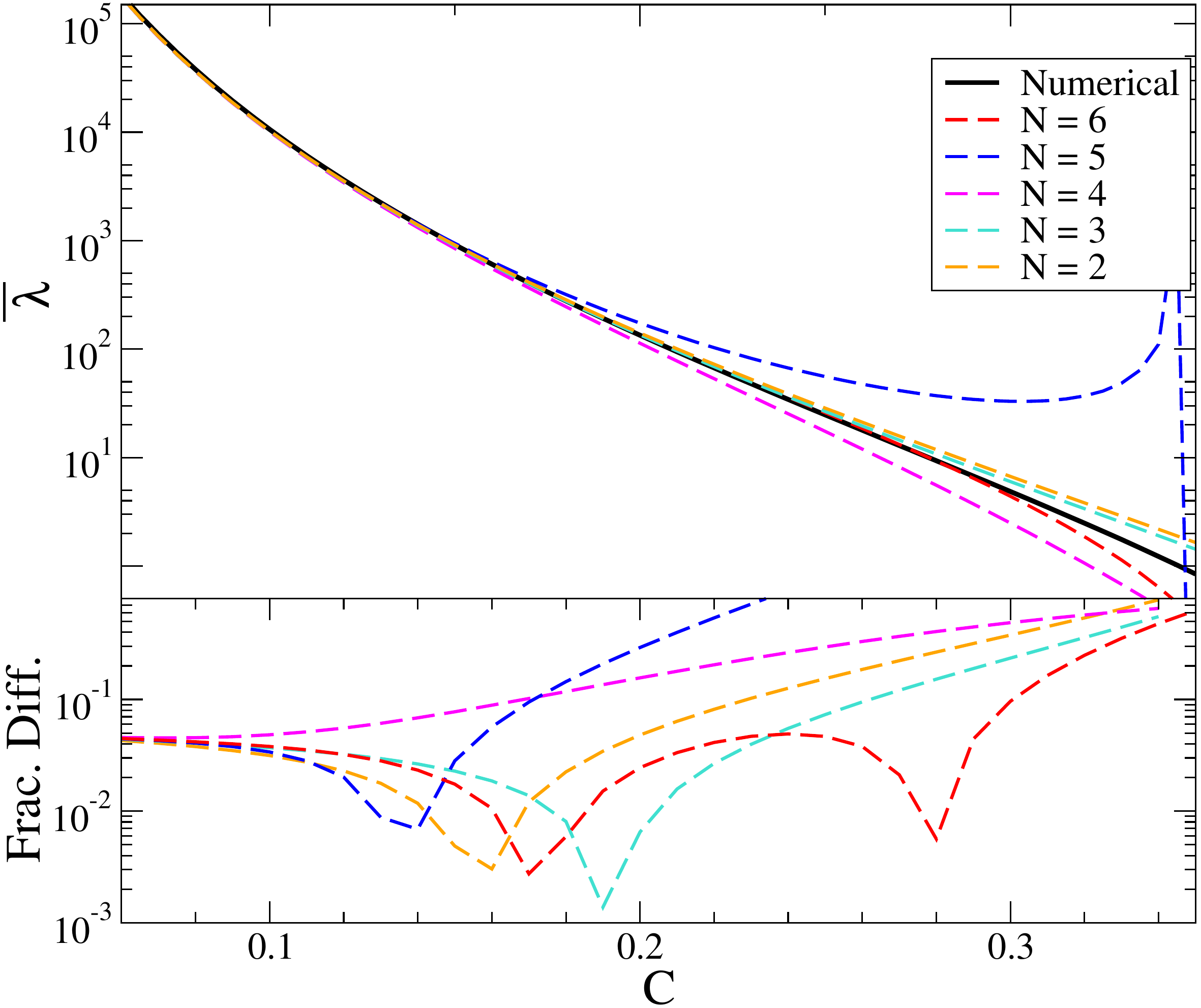}
\caption{Similar to Fig.~\ref{fig:IC converge} but for the Love-C relation. Observe that the series does not converge in this case and the 3rd order expansion gives the most accurate result.}
\label{fig:Love converge}
\end{figure}

\begin{figure}[htp]
\includegraphics[width=8.5cm]{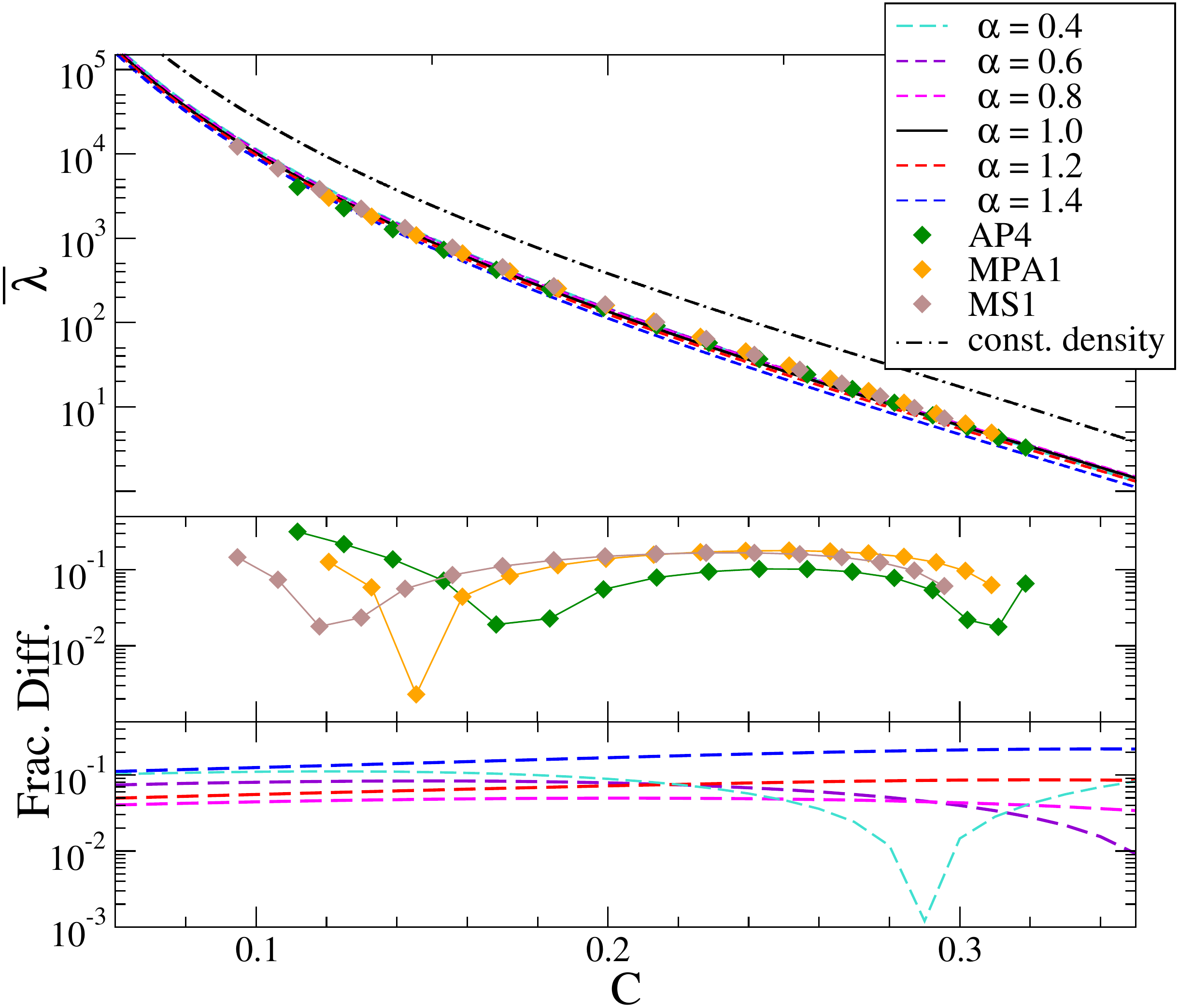}
\caption{
Similar to Fig.~\ref{fig:I-C} but for the Love-C relation. The analytic relations for the modifiend Tolman model have been kept up to 3rd order in $\mathcal{C}$. Observe that the EoS-variation  is kept under $\mathcal{O}(10\%)$ for both modified Tolman VII and realistic EoSs, in agreement with that for realistic EoSs in Fig.~15 of~\cite{YAGI20171}}
\label{fig:Love-C}
\end{figure}

The top panel of Fig.~\ref{fig:Love-C} presents the analytic Love-C relations for the modified Tolman solution, together with those for the realistic EoSs and the one found in~\cite{PhysRevD.91.044017} for constant density stars. Observe that similar to Fig.~\ref{fig:I-C}, the analytic relation for the modified Tolman solution can accurately describe the results for the realistic EoSs while that for constant density stars fails to do so. The bottom panel shows the fractional difference in the Love-C relations between the modified and original Tolman solutions. Observe that the amount of the variation in $\alpha$ is of order 10\%, consistent with the EoS-variation in the Love-C relation for realistic EoSs in the middle panel (see also~\cite{YAGI20171}).

Similar to the I-C case, we can investigate further the amount of the $\alpha$ (or EoS) variation in the Love-C relation by expanding the analytic solution about the original Tolman solution. To achieve this, we first expand $\bar \lambda$ as
\begin{equation}
\bar \lambda \approx \sum_{n=0}^N \bar\lambda_n \mathcal{C}^n.
\end{equation}
We then expand further each coefficient $\bar \lambda_n$ about $\alpha = 1$ and find
\begin{align*}
&\bar{\lambda}_0 = 0.204 [ 1 - 0.186 (\alpha - 1 ) + \mathcal{O}(\alpha - 1)^2 ]\\
&\bar{\lambda}_1 = -1.274 [ 1 - 0.090 (\alpha - 1 ) + \mathcal{O}(\alpha - 1)^2 ]\\
&\bar{\lambda}_2 = 2.749 [ 1 + 0.082 (\alpha - 1 ) + \mathcal{O}(\alpha - 1)^2 ]\\
&\bar{\lambda}_3 = -2.281 [ 1 + 0.420 (\alpha - 1 ) + \mathcal{O}(\alpha - 1)^2 ]\\
&\bar{\lambda}_4 = 3.199 [ 1 + 0.013 (\alpha - 1 ) + \mathcal{O}(\alpha - 1)^2 ]\\
&\bar{\lambda}_5 = -13.001 [ 1 - 0.108 (\alpha - 1 ) + \mathcal{O}(\alpha - 1)^2 ]\\
&\bar{\lambda}_6 = 19.282 [ 1 + 0.161 (\alpha - 1 ) + \mathcal{O}(\alpha - 1)^2 ].
\end{align*}
This expansion shows that the fractional $\alpha$ variation is only of order 10\% or less for most of the coefficients, which again is consistent with the amount of EoS variation in the Love-C relation in Fig.~\ref{fig:Love-C}. This gives us an analytic explanation for the amount of EoS variation in the relation.

\section{\label{sec:4}I-Love relation}

Now that we have calculated the moment of inertia and tidal deformability, we can construct semi-analytic relations between these two quantities based on the modified Tolman VII solutions.

One can derive the I-Love relation ($\bar I$ as a function of $\bar \lambda$) semi-analytically based on the analytic I-C and Love-C relations by parametrically plotting $\bar I$ and $\bar\lambda$. Namely, we choose one $\mathcal{C}$ and plot a point in the I-Love plane using the above two analytic relations. We then change $\mathcal{C}$ and repeat the procedure to find the I-Love relation. This is semi-analytic in the sense that we are not providing a closed form expression of $\bar I$ in terms of $\bar \lambda$. 

Regarding the Love-C relation, we found that the expansion to 6th order in $\mathcal C$ gives us a more accurate semi-analytic I-Love relation than that with the 3rd-order expansion, which is counter-intuitive given that the 6th-order Love-C relation is less accurate than the 3rd-order one. This behavior arises because the amount of inaccuracy in the 6th-order Love-C relation partially cancels with that in the I-C relation, producing a more accurate I-Love relation. For this reason, we will use the 6th order Love-C relation for constructing the I-Love relation.

The top panel of Fig.~\ref{fig:I-Love} presents such semi-analytic I-Love relations for the modified Tolman VII model. In most regime, observe that the $\alpha$-variation shown in the bottom panel is of $\mathcal{O}(1\%)$ at most. This reproduces the amount of the EoS-variation in the I-Love relations for realistic NSs shown in the middle panel and also found in~\cite{Yagi365,Yagi:2013awa,YAGI20171}. The $\alpha$ variation becomes slightly larger when $\bar I$ and $\bar \lambda$ are smaller, though this is due to the fact that the sixth order Love-C relation becoming less accurate in the large $\mathcal{C}$ regime.

\begin{figure}[htp]
\includegraphics[width=8.5cm]{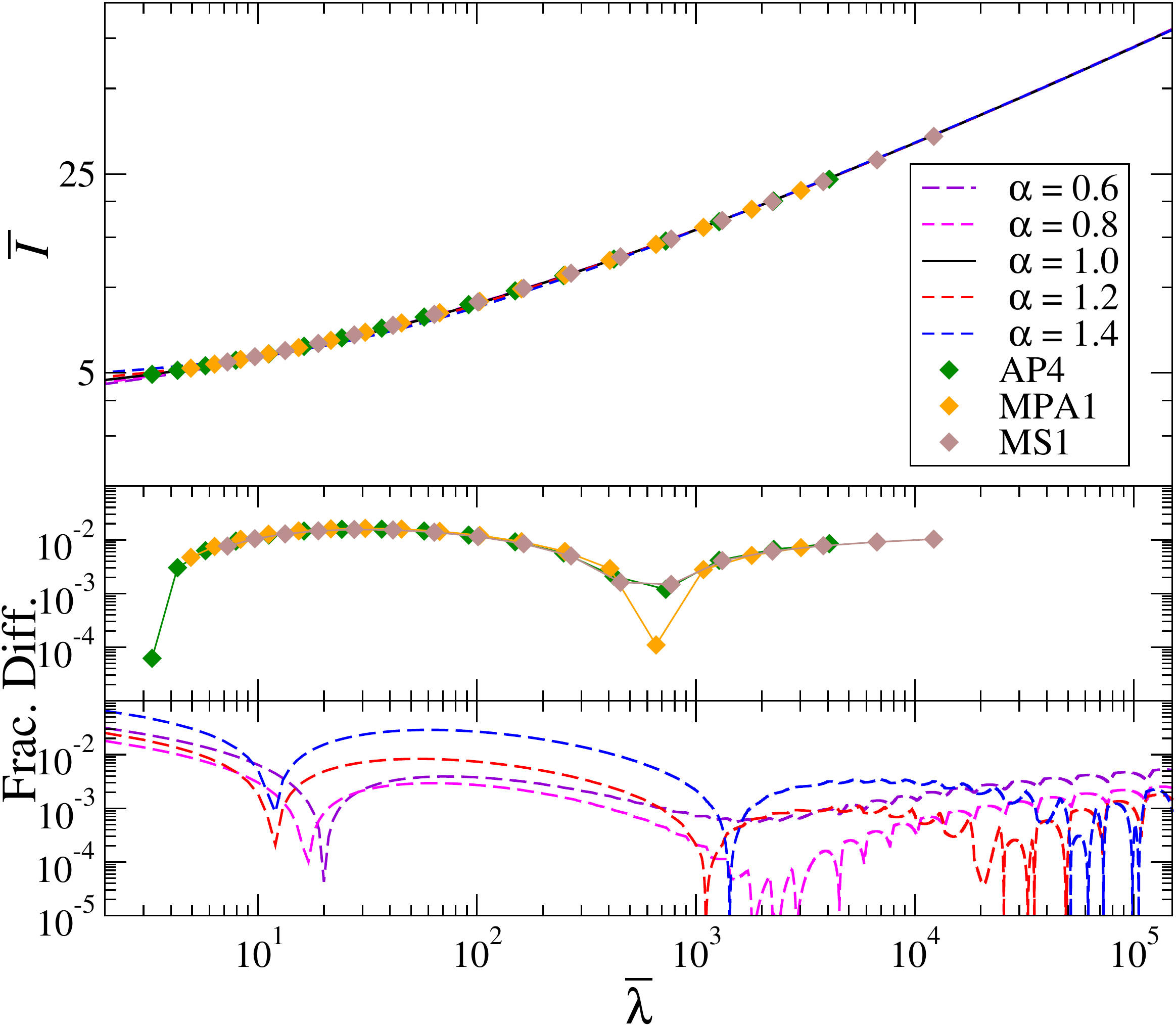}
\caption{(Top) 
Similar to Fig.~\ref{fig:I-C} but for I-Love relation. The semi-analytic modified TolmanVII model is constructed by keeping the analytic I-C and Love-C relations up to 6th order in $\mathcal{C}$. Observe that the EoS-variation  is kept around 1\% for both realistic EoSs and modified Tolman VII model. }
\label{fig:I-Love}
\end{figure}

One can also obtain an approximate but completely analytic I-Love relation by first inverting the Love-C relation for $\mathcal{C}$ and substituting it to the I-C relation. For inverting the Love-C relation, one can find a series expanded solution of the form
\begin{equation}
\mathcal C = \sum_{n=0} \frac{a_n}{\bar{\lambda}^{n/5}},
\end{equation}
with the coefficients $a_n$ that is a function of $\alpha$. The analytic I-Love relation obtained in this way, however, is not very accurate, at least up to the (sixth) order that we have worked on. We found that the variation in $\alpha$ can be $\mathcal{O}(10\%)$ or larger, and one may need to increase the order of the expansion to find a more accurate relation, which we leave for future work.

One can improve the semi-analytic model further by resumming the series-expanded I-C and Love-C (or the $y_R$ expression to be more precise) relations using Pad\'e approximant. Such Pad\'e expressions for the modified Tolman solutions are lengthy and not very useful, while we managed to derive a simple form for the original Tolman solution. The results are given in Eqs.~\eqref{Pade I first}--\eqref{Pade Love first}, and the relations are show in Fig.~\ref{fig:PadeTolman}. For the Pad\'e resummation, we found that the 3rd order expansion of $y(\mathcal{C})$ provides a more accurate, semi-analytic I-Love relation (to be discussed in the next paragraph) than the 6th order expansion. This is why the  Pad\'e resummation is of order (3,3) for the I-C relation, while is of order (2,2)  for the Love-C relation.

The top panel of Fig.~\ref{fig:TolILove} presents the semi-analytic I-Love relation for the original Tolman solution using the Pad\'e expressions for the I-C and Love-C relations. For comparison, we also present the Pad\'e-resummed relation for constant density stars~\cite{PhysRevD.91.044017} and realistic NSs using all 11 EoSs considered in Fig.~\ref{fig:alpha-C}. The middle (bottom) panel shows the relative fractional difference between the relation with each EoS and the constant-density (Pad\'e-resummed original Tolman) model. Observe that the original Tolman relation always agrees with the realistic models within an error of 1\% and gives more accurate description than the constant density case especially in the large compactness (small $\bar \lambda$) regime. Comparing the bottom panel of Fig.~\ref{fig:TolILove} with the middle panel of Fig.~\ref{fig:I-Love}, one sees that the former slightly has a smaller fractional error, and thus the Pad\'e resummed I-Love relation indeed proves us a more accurate result than the series-expanded one. We do not show the numerical fit for the I-Love curve found in~\cite{Yagi365} since it is indistinguishable from other curves and the fractional differences are around 1\%.

\begin{figure}[htp]
\includegraphics[width=8.5cm]{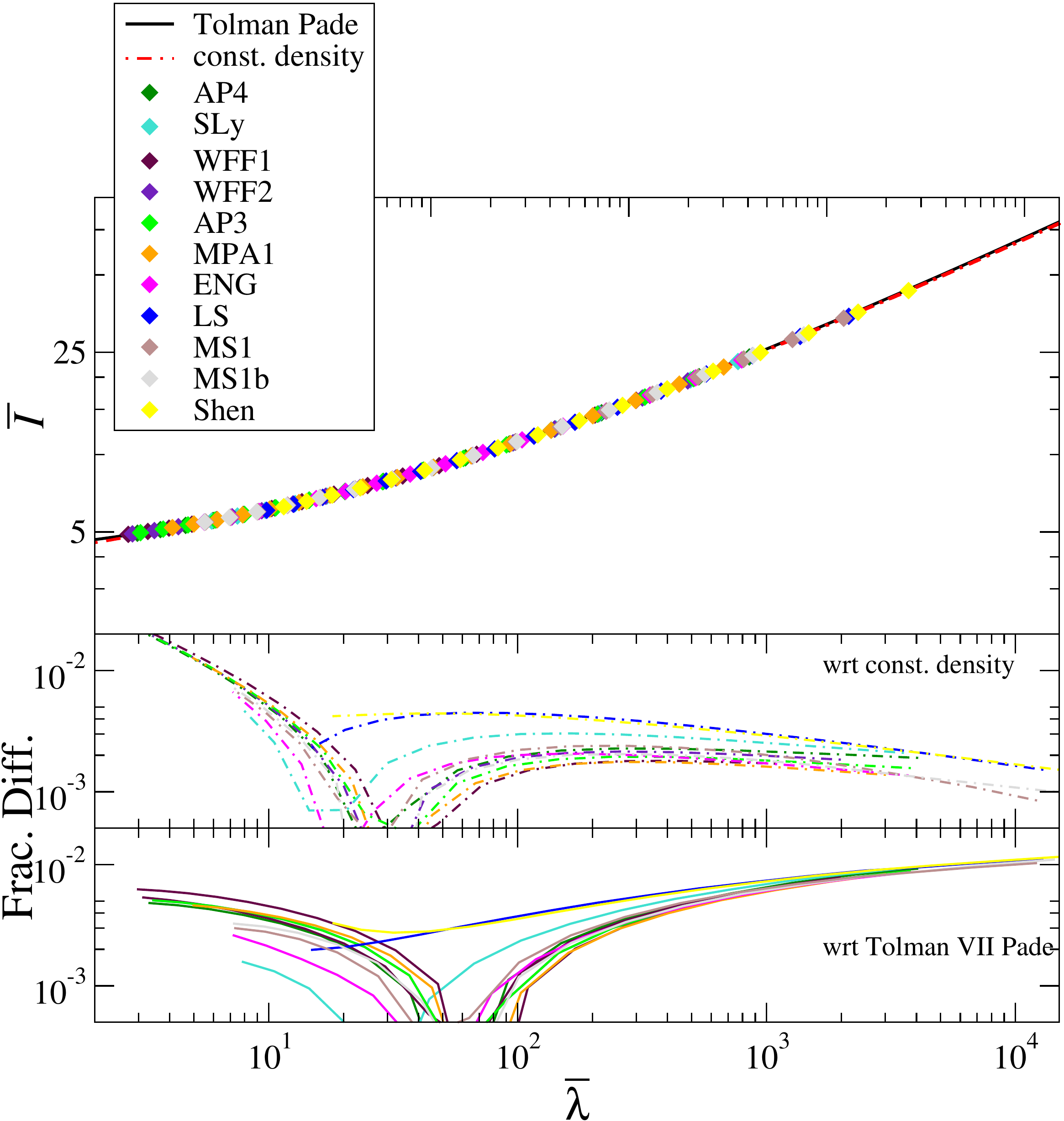}
\caption{
(Top) Semi-analytic, Pad\'e-resummed I-Love relation for the original Tolman VII solution, together with the analytic relation for constant density stars~\cite{PhysRevD.91.044017} and numerical results for realistic EoSs. (Middle) The relative fractional difference of the relations for the realistic EoSs from the analytic one for constant density stars. (Bottom) Similar to the middle panel but for the Tolman VII relation.
Observe that the Tolman VII model maintains universality with an EoS-variation of less than 1\% for the whole range of $\bar \lambda$. }
\label{fig:TolILove}
\end{figure}
\section{Possible Origin of the Universality}
\label{sec:origin}

\begin{figure}[htp]
\includegraphics[width=8.5cm]{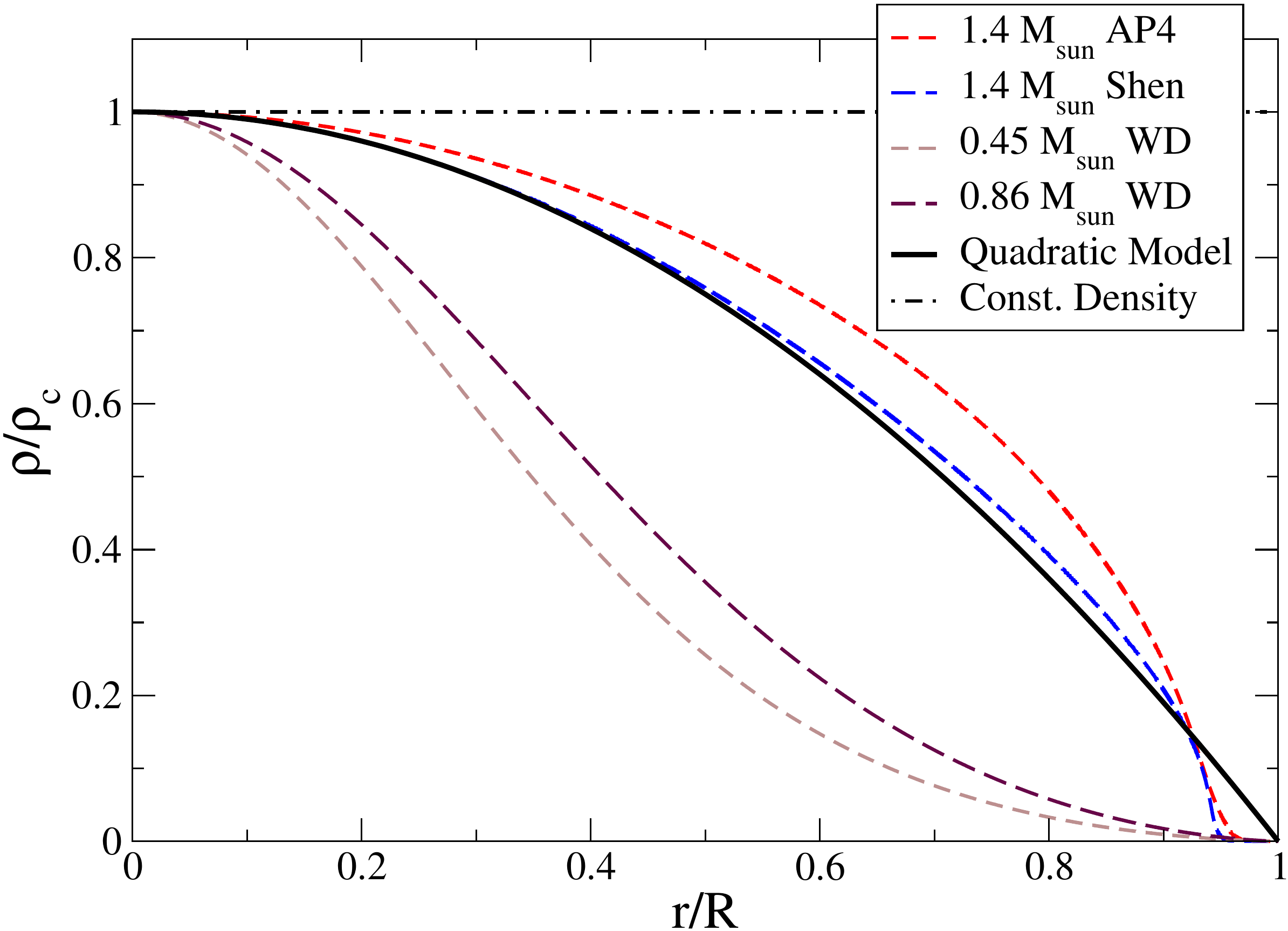}
\caption{The normalized energy density profiles of NSs and white dwarfs, together with quadratic (original Tolman VII solution) and constant density models. The fact that realistic NS profiles approximately follow the quadratic function in Eq.~\eqref{eq:rho-Tol} (whereas constant density stars and white dwarfs do not) can be interpreted as a possible origin of the I-C and Love-C universality.}
\label{fig:densityprofile}
\end{figure}

We now try to connect our analytic calculations to possible origins of some of the universal relations. Regarding the I-Love-Q relations, Ref.~\cite{Yagi:2014qua} showed that the universality is related to approximate self-similarity of isodensity contours inside NSs, and this self-similarity becomes exact for constant density stars. Similarly, Ref.~\cite{Sham_2015} demonstrated that the origin of the universality can be attributed to the fact that realistic EoSs for NSs are ``similar'' to constant density EoSs, which is consistent with the findings in~\cite{Yagi:2014qua}.  Unfortunately, these explanations do not apply to the I-C and Love-C relations since NSs and constant density stars do not share the universality.

Here, we aim to obtain some insight on the possible origin of the I-C and Love-C universal relations based on our analytic calculations presented in the previous sections. The original Tolman VII solution is a two-parameter model characterized by the stellar mass and radius and thus can explain any mass-radius relations for NSs. Yet, we found that there is a unique I-C and Love-C relations for the Tolman VII solution that agree well with those found numerically with realistic EoSs. The underlying assumption in the Tolman VII solution is that the energy density profile inside a star follows a quadratic function in the radial coordinate (Eq.~\eqref{eq:rho-Tol}). All of this suggest that one can attribute the origin of the I-C and Love-C universality to the fact that the energy density profile for NSs roughly follows a quadratic function.

Figure~\ref{fig:densityprofile} presents the energy density profiles for realistic NSs, together with realistic white dwarfs, Tolman VII quadratic model and constant density stars. Observe that the NS profiles can be approximated well with the quadratic model, while white dwarfs and constant density stars have different profiles. Universal relations for white dwarfs have been studied e.g. in~\cite{ilq_wd,hotwd,Taylor:2019hle}, which showed that they were quite different from the NS ones. The modified Tolman VII model that we also considered in this paper accounts for the difference between the realistic NS profile and the quadratic one. As shown in the previous sections, we managed to recover the $\mathcal{O}(10\%)$ EoS-variation analytically. Figure~\ref{fig:densityprofile} together with these analytic calculations support our claim that one possible origin of the universality for the I-C and Love-C relations is due to the fact that realistic NSs follow more or less the quadratic energy density profile and the deviation can explain the amount of the EoS-variation in the relations.

\section{\label{sec:Conclusion}Conclusions and Discussions}

In this paper, we derived analytic I-C, Love-C and semi-analytic I-Love relations based on the modified Tolman VII model. We used series expansion in compactness to solve differential equations and found approximate expressions for the moment of inertia and tidal deformability. 
Varying the modified Tolman VII parameter $\alpha$ that enters in the energy density profile, we analytically  showed the 10\%EoS-variation in the I-C and Love-C relations, and the 1\% EoS-variation  in the I-Love relation that are consistent with the amount found numerically~\cite{YAGI20171}}. Our analytic relations more accurately describe the relations for realistic NSs than the constant density model~\cite{PhysRevD.91.044017} (more appropriate for quark stars~\cite{YAGI20171}), especially for the I-C and Love-C relations. Our analytic relations are comparable to the the fitted relations in their accuracy~\cite{YAGI20171}, and the former is based on a theoretical founding while the latter is phenomenological. Based on these analytic findings, we pointed out that a possible origin of the I-C and Love-C relations may be due to the energy density profile for realistic NSs approximately following a quadratic function. This can explain why the constant density stars deviates from the NS branch in these relations. Such an origin is different from that for the universality in the I-Love-Q relations discussed in~\cite{Yagi:2014qua,Sham_2015}.

 Let us comment on a few possible directions for future work. One possibility includes improving the analytic results presented here. For example, one could try a different expansion in $\xi$ or $\mathcal{C}$ for the Love-C relation, such as that in~\cite{Petroff:2007tz} for constant density stars. One could then try to construct a more accurate inversion of the relation, so that an accurate, analytic I-Love relation can be derived. Another avenue includes deriving similar, analytic expressions for multipole moments, such as quadrupole and octupole, so that one could construct, analytic, I-Love-Q relations~\cite{Yagi365,Yagi:2013awa} and ``no-hair'' relations~\cite{Stein:2013ofa,Yagi:2014bxa,Majumder:2015kfa} for neutron stars. One could also derive analytic expressions for multipolar tidal deformabilities and construct analytic multipole Love relations~\cite{Yagi:2013sva}.
Such studies may further extend our knowledge of the origin of these universal relations.

\acknowledgments
We thank Andy Taylor for providing us the energy density profile for realistic white dwarfs.
N.J. and K.Y. acknowledge support from the Ed Owens Fund. K.Y. would like to also acknowledge support from NSF Award PHY-1806776, a Sloan Foundation Research Fellowship, the COST Action GWverse CA16104 and JSPS KAKENHI Grants No. JP17H06358.

\bibliography{ref}

\begin{thebibliography}{65}%
\makeatletter
\providecommand \@ifxundefined [1]{%
 \@ifx{#1\undefined}
}%
\providecommand \@ifnum [1]{%
 \ifnum #1\expandafter \@firstoftwo
 \else \expandafter \@secondoftwo
 \fi
}%
\providecommand \@ifx [1]{%
 \ifx #1\expandafter \@firstoftwo
 \else \expandafter \@secondoftwo
 \fi
}%
\providecommand \natexlab [1]{#1}%
\providecommand \enquote  [1]{``#1''}%
\providecommand \bibnamefont  [1]{#1}%
\providecommand \bibfnamefont [1]{#1}%
\providecommand \citenamefont [1]{#1}%
\providecommand \href@noop [0]{\@secondoftwo}%
\providecommand \href [0]{\begingroup \@sanitize@url \@href}%
\providecommand \@href[1]{\@@startlink{#1}\@@href}%
\providecommand \@@href[1]{\endgroup#1\@@endlink}%
\providecommand \@sanitize@url [0]{\catcode `\\12\catcode `\$12\catcode
  `\&12\catcode `\#12\catcode `\^12\catcode `\_12\catcode `\%12\relax}%
\providecommand \@@startlink[1]{}%
\providecommand \@@endlink[0]{}%
\providecommand \url  [0]{\begingroup\@sanitize@url \@url }%
\providecommand \@url [1]{\endgroup\@href {#1}{\urlprefix }}%
\providecommand \urlprefix  [0]{URL }%
\providecommand \Eprint [0]{\href }%
\providecommand \doibase [0]{http://dx.doi.org/}%
\providecommand \selectlanguage [0]{\@gobble}%
\providecommand \bibinfo  [0]{\@secondoftwo}%
\providecommand \bibfield  [0]{\@secondoftwo}%
\providecommand \translation [1]{[#1]}%
\providecommand \BibitemOpen [0]{}%
\providecommand \bibitemStop [0]{}%
\providecommand \bibitemNoStop [0]{.\EOS\space}%
\providecommand \EOS [0]{\spacefactor3000\relax}%
\providecommand \BibitemShut  [1]{\csname bibitem#1\endcsname}%
\let\auto@bib@innerbib\@empty
\bibitem [{\citenamefont {Lattimer}\ and\ \citenamefont
  {Prakash}(2001)}]{Lattimer:2000nx}%
  \BibitemOpen
  \bibfield  {author} {\bibinfo {author} {\bibfnamefont {J.}~\bibnamefont
  {Lattimer}}\ and\ \bibinfo {author} {\bibfnamefont {M.}~\bibnamefont
  {Prakash}},\ }\href {\doibase 10.1086/319702} {\bibfield  {journal} {\bibinfo
   {journal} {Astrophys.J.}\ }\textbf {\bibinfo {volume} {550}},\ \bibinfo
  {pages} {426} (\bibinfo {year} {2001})},\ \Eprint
  {http://arxiv.org/abs/astro-ph/0002232} {arXiv:astro-ph/0002232} \BibitemShut
  {NoStop}%
\bibitem [{\citenamefont {Guver}\ and\ \citenamefont {Ozel}(2013)}]{guver}%
  \BibitemOpen
  \bibfield  {author} {\bibinfo {author} {\bibfnamefont {T.}~\bibnamefont
  {Guver}}\ and\ \bibinfo {author} {\bibfnamefont {F.}~\bibnamefont {Ozel}},\
  }\href {\doibase 10.1088/2041-8205/765/1/L1} {\bibfield  {journal} {\bibinfo
  {journal} {Astrophys. J.}\ }\textbf {\bibinfo {volume} {765}},\ \bibinfo
  {pages} {L1} (\bibinfo {year} {2013})},\ \Eprint
  {http://arxiv.org/abs/1301.0831} {arXiv:1301.0831 [astro-ph.HE]} \BibitemShut
  {NoStop}%
\bibitem [{\citenamefont {Steiner}\ \emph {et~al.}(2010)\citenamefont
  {Steiner}, \citenamefont {Lattimer},\ and\ \citenamefont
  {Brown}}]{steiner-lattimer-brown}%
  \BibitemOpen
  \bibfield  {author} {\bibinfo {author} {\bibfnamefont {A.~W.}\ \bibnamefont
  {Steiner}}, \bibinfo {author} {\bibfnamefont {J.~M.}\ \bibnamefont
  {Lattimer}}, \ and\ \bibinfo {author} {\bibfnamefont {E.~F.}\ \bibnamefont
  {Brown}},\ }\href {\doibase 10.1088/0004-637X/722/1/33} {\bibfield  {journal}
  {\bibinfo  {journal} {Astrophys.J.}\ }\textbf {\bibinfo {volume} {722}},\
  \bibinfo {pages} {33} (\bibinfo {year} {2010})}\BibitemShut {NoStop}%
\bibitem [{\citenamefont {Lattimer}\ and\ \citenamefont
  {Steiner}(2014)}]{Lattimer2014}%
  \BibitemOpen
  \bibfield  {author} {\bibinfo {author} {\bibfnamefont {J.~M.}\ \bibnamefont
  {Lattimer}}\ and\ \bibinfo {author} {\bibfnamefont {A.~W.}\ \bibnamefont
  {Steiner}},\ }\href {\doibase 10.1140/epja/i2014-14040-y} {\bibfield
  {journal} {\bibinfo  {journal} {The European Physical Journal A}\ }\textbf
  {\bibinfo {volume} {50}} (\bibinfo {year} {2014}),\
  10.1140/epja/i2014-14040-y}\BibitemShut {NoStop}%
\bibitem [{\citenamefont {Ozel}\ \emph {et~al.}(2016)\citenamefont {Ozel},
  \citenamefont {Psaltis}, \citenamefont {Guver}, \citenamefont {Baym},
  \citenamefont {Heinke},\ and\ \citenamefont {Guillot}}]{Ozel:2015fia}%
  \BibitemOpen
  \bibfield  {author} {\bibinfo {author} {\bibfnamefont {F.}~\bibnamefont
  {Ozel}}, \bibinfo {author} {\bibfnamefont {D.}~\bibnamefont {Psaltis}},
  \bibinfo {author} {\bibfnamefont {T.}~\bibnamefont {Guver}}, \bibinfo
  {author} {\bibfnamefont {G.}~\bibnamefont {Baym}}, \bibinfo {author}
  {\bibfnamefont {C.}~\bibnamefont {Heinke}}, \ and\ \bibinfo {author}
  {\bibfnamefont {S.}~\bibnamefont {Guillot}},\ }\href {\doibase
  10.3847/0004-637X/820/1/28} {\bibfield  {journal} {\bibinfo  {journal}
  {Astrophys. J.}\ }\textbf {\bibinfo {volume} {820}},\ \bibinfo {pages} {28}
  (\bibinfo {year} {2016})},\ \Eprint {http://arxiv.org/abs/1505.05155}
  {arXiv:1505.05155 [astro-ph.HE]} \BibitemShut {NoStop}%
\bibitem [{\citenamefont {Ozel}\ and\ \citenamefont
  {Freire}(2016)}]{Ozel:2016oaf}%
  \BibitemOpen
  \bibfield  {author} {\bibinfo {author} {\bibfnamefont {F.}~\bibnamefont
  {Ozel}}\ and\ \bibinfo {author} {\bibfnamefont {P.}~\bibnamefont {Freire}},\
  }\href {\doibase 10.1146/annurev-astro-081915-023322} {\bibfield  {journal}
  {\bibinfo  {journal} {Ann. Rev. Astron. Astrophys.}\ }\textbf {\bibinfo
  {volume} {54}},\ \bibinfo {pages} {401} (\bibinfo {year} {2016})},\ \Eprint
  {http://arxiv.org/abs/1603.02698} {arXiv:1603.02698 [astro-ph.HE]}
  \BibitemShut {NoStop}%
\bibitem [{\citenamefont {Steiner}\ \emph {et~al.}(2018)\citenamefont
  {Steiner}, \citenamefont {Heinke}, \citenamefont {Bogdanov}, \citenamefont
  {Li}, \citenamefont {Ho}, \citenamefont {Bahramian},\ and\ \citenamefont
  {Han}}]{Steiner:2017vmg}%
  \BibitemOpen
  \bibfield  {author} {\bibinfo {author} {\bibfnamefont {A.~W.}\ \bibnamefont
  {Steiner}}, \bibinfo {author} {\bibfnamefont {C.~O.}\ \bibnamefont {Heinke}},
  \bibinfo {author} {\bibfnamefont {S.}~\bibnamefont {Bogdanov}}, \bibinfo
  {author} {\bibfnamefont {C.}~\bibnamefont {Li}}, \bibinfo {author}
  {\bibfnamefont {W.~C.~G.}\ \bibnamefont {Ho}}, \bibinfo {author}
  {\bibfnamefont {A.}~\bibnamefont {Bahramian}}, \ and\ \bibinfo {author}
  {\bibfnamefont {S.}~\bibnamefont {Han}},\ }\href {\doibase
  10.1093/mnras/sty215} {\bibfield  {journal} {\bibinfo  {journal} {Mon. Not.
  Roy. Astron. Soc.}\ }\textbf {\bibinfo {volume} {476}},\ \bibinfo {pages}
  {421} (\bibinfo {year} {2018})},\ \Eprint {http://arxiv.org/abs/1709.05013}
  {arXiv:1709.05013 [astro-ph.HE]} \BibitemShut {NoStop}%
\bibitem [{\citenamefont {Riley}\ \emph {et~al.}(2019)\citenamefont {Riley},
  \citenamefont {Watts}, \citenamefont {Bogdanov}, \citenamefont {Ray},
  \citenamefont {Ludlam}, \citenamefont {Guillot}, \citenamefont {Arzoumanian},
  \citenamefont {Baker}, \citenamefont {Bilous}, \citenamefont {Chakrabarty},
  \citenamefont {Gendreau}, \citenamefont {Harding}, \citenamefont {Ho},
  \citenamefont {Lattimer}, \citenamefont {Morsink},\ and\ \citenamefont
  {Strohmayer}}]{Riley_2019}%
  \BibitemOpen
  \bibfield  {author} {\bibinfo {author} {\bibfnamefont {T.~E.}\ \bibnamefont
  {Riley}}, \bibinfo {author} {\bibfnamefont {A.~L.}\ \bibnamefont {Watts}},
  \bibinfo {author} {\bibfnamefont {S.}~\bibnamefont {Bogdanov}}, \bibinfo
  {author} {\bibfnamefont {P.~S.}\ \bibnamefont {Ray}}, \bibinfo {author}
  {\bibfnamefont {R.~M.}\ \bibnamefont {Ludlam}}, \bibinfo {author}
  {\bibfnamefont {S.}~\bibnamefont {Guillot}}, \bibinfo {author} {\bibfnamefont
  {Z.}~\bibnamefont {Arzoumanian}}, \bibinfo {author} {\bibfnamefont {C.~L.}\
  \bibnamefont {Baker}}, \bibinfo {author} {\bibfnamefont {A.~V.}\ \bibnamefont
  {Bilous}}, \bibinfo {author} {\bibfnamefont {D.}~\bibnamefont {Chakrabarty}},
  \bibinfo {author} {\bibfnamefont {K.~C.}\ \bibnamefont {Gendreau}}, \bibinfo
  {author} {\bibfnamefont {A.~K.}\ \bibnamefont {Harding}}, \bibinfo {author}
  {\bibfnamefont {W.~C.~G.}\ \bibnamefont {Ho}}, \bibinfo {author}
  {\bibfnamefont {J.~M.}\ \bibnamefont {Lattimer}}, \bibinfo {author}
  {\bibfnamefont {S.~M.}\ \bibnamefont {Morsink}}, \ and\ \bibinfo {author}
  {\bibfnamefont {T.~E.}\ \bibnamefont {Strohmayer}},\ }\href {\doibase
  10.3847/2041-8213/ab481c} {\bibfield  {journal} {\bibinfo  {journal} {The
  Astrophysical Journal}\ }\textbf {\bibinfo {volume} {887}},\ \bibinfo {pages}
  {L21} (\bibinfo {year} {2019})}\BibitemShut {NoStop}%
\bibitem [{\citenamefont {Miller}\ \emph {et~al.}(2019)\citenamefont {Miller},
  \citenamefont {Lamb}, \citenamefont {Dittmann}, \citenamefont {Bogdanov},
  \citenamefont {Arzoumanian}, \citenamefont {Gendreau}, \citenamefont
  {Guillot}, \citenamefont {Harding}, \citenamefont {Ho}, \citenamefont
  {Lattimer}, \citenamefont {Ludlam}, \citenamefont {Mahmoodifar},
  \citenamefont {Morsink}, \citenamefont {Ray}, \citenamefont {Strohmayer},
  \citenamefont {Wood}, \citenamefont {Enoto}, \citenamefont {Foster},
  \citenamefont {Okajima}, \citenamefont {Prigozhin},\ and\ \citenamefont
  {Soong}}]{Miller_2019}%
  \BibitemOpen
  \bibfield  {author} {\bibinfo {author} {\bibfnamefont {M.~C.}\ \bibnamefont
  {Miller}}, \bibinfo {author} {\bibfnamefont {F.~K.}\ \bibnamefont {Lamb}},
  \bibinfo {author} {\bibfnamefont {A.~J.}\ \bibnamefont {Dittmann}}, \bibinfo
  {author} {\bibfnamefont {S.}~\bibnamefont {Bogdanov}}, \bibinfo {author}
  {\bibfnamefont {Z.}~\bibnamefont {Arzoumanian}}, \bibinfo {author}
  {\bibfnamefont {K.~C.}\ \bibnamefont {Gendreau}}, \bibinfo {author}
  {\bibfnamefont {S.}~\bibnamefont {Guillot}}, \bibinfo {author} {\bibfnamefont
  {A.~K.}\ \bibnamefont {Harding}}, \bibinfo {author} {\bibfnamefont
  {W.~C.~G.}\ \bibnamefont {Ho}}, \bibinfo {author} {\bibfnamefont {J.~M.}\
  \bibnamefont {Lattimer}}, \bibinfo {author} {\bibfnamefont {R.~M.}\
  \bibnamefont {Ludlam}}, \bibinfo {author} {\bibfnamefont {S.}~\bibnamefont
  {Mahmoodifar}}, \bibinfo {author} {\bibfnamefont {S.~M.}\ \bibnamefont
  {Morsink}}, \bibinfo {author} {\bibfnamefont {P.~S.}\ \bibnamefont {Ray}},
  \bibinfo {author} {\bibfnamefont {T.~E.}\ \bibnamefont {Strohmayer}},
  \bibinfo {author} {\bibfnamefont {K.~S.}\ \bibnamefont {Wood}}, \bibinfo
  {author} {\bibfnamefont {T.}~\bibnamefont {Enoto}}, \bibinfo {author}
  {\bibfnamefont {R.}~\bibnamefont {Foster}}, \bibinfo {author} {\bibfnamefont
  {T.}~\bibnamefont {Okajima}}, \bibinfo {author} {\bibfnamefont
  {G.}~\bibnamefont {Prigozhin}}, \ and\ \bibinfo {author} {\bibfnamefont
  {Y.}~\bibnamefont {Soong}},\ }\href {\doibase 10.3847/2041-8213/ab50c5}
  {\bibfield  {journal} {\bibinfo  {journal} {The Astrophysical Journal}\
  }\textbf {\bibinfo {volume} {887}},\ \bibinfo {pages} {L24} (\bibinfo {year}
  {2019})}\BibitemShut {NoStop}%
\bibitem [{\citenamefont {Bogdanov}\ \emph
  {et~al.}(2019{\natexlab{a}})\citenamefont {Bogdanov}, \citenamefont
  {Guillot}, \citenamefont {Ray}, \citenamefont {Wolff}, \citenamefont
  {Chakrabarty}, \citenamefont {Ho}, \citenamefont {Kerr}, \citenamefont
  {Lamb}, \citenamefont {Lommen}, \citenamefont {Ludlam}, \citenamefont
  {Milburn}, \citenamefont {Montano}, \citenamefont {Miller}, \citenamefont
  {Bauböck}, \citenamefont {Özel}, \citenamefont {Psaltis}, \citenamefont
  {Remillard}, \citenamefont {Riley}, \citenamefont {Steiner}, \citenamefont
  {Strohmayer}, \citenamefont {Watts}, \citenamefont {Wood}, \citenamefont
  {Zeldes}, \citenamefont {Enoto}, \citenamefont {Okajima}, \citenamefont
  {Kellogg}, \citenamefont {Baker}, \citenamefont {Markwardt}, \citenamefont
  {Arzoumanian},\ and\ \citenamefont {Gendreau}}]{Bogdanov_2019}%
  \BibitemOpen
  \bibfield  {author} {\bibinfo {author} {\bibfnamefont {S.}~\bibnamefont
  {Bogdanov}}, \bibinfo {author} {\bibfnamefont {S.}~\bibnamefont {Guillot}},
  \bibinfo {author} {\bibfnamefont {P.~S.}\ \bibnamefont {Ray}}, \bibinfo
  {author} {\bibfnamefont {M.~T.}\ \bibnamefont {Wolff}}, \bibinfo {author}
  {\bibfnamefont {D.}~\bibnamefont {Chakrabarty}}, \bibinfo {author}
  {\bibfnamefont {W.~C.~G.}\ \bibnamefont {Ho}}, \bibinfo {author}
  {\bibfnamefont {M.}~\bibnamefont {Kerr}}, \bibinfo {author} {\bibfnamefont
  {F.~K.}\ \bibnamefont {Lamb}}, \bibinfo {author} {\bibfnamefont
  {A.}~\bibnamefont {Lommen}}, \bibinfo {author} {\bibfnamefont {R.~M.}\
  \bibnamefont {Ludlam}}, \bibinfo {author} {\bibfnamefont {R.}~\bibnamefont
  {Milburn}}, \bibinfo {author} {\bibfnamefont {S.}~\bibnamefont {Montano}},
  \bibinfo {author} {\bibfnamefont {M.~C.}\ \bibnamefont {Miller}}, \bibinfo
  {author} {\bibfnamefont {M.}~\bibnamefont {Bauböck}}, \bibinfo {author}
  {\bibfnamefont {F.}~\bibnamefont {Özel}}, \bibinfo {author} {\bibfnamefont
  {D.}~\bibnamefont {Psaltis}}, \bibinfo {author} {\bibfnamefont {R.~A.}\
  \bibnamefont {Remillard}}, \bibinfo {author} {\bibfnamefont {T.~E.}\
  \bibnamefont {Riley}}, \bibinfo {author} {\bibfnamefont {J.~F.}\ \bibnamefont
  {Steiner}}, \bibinfo {author} {\bibfnamefont {T.~E.}\ \bibnamefont
  {Strohmayer}}, \bibinfo {author} {\bibfnamefont {A.~L.}\ \bibnamefont
  {Watts}}, \bibinfo {author} {\bibfnamefont {K.~S.}\ \bibnamefont {Wood}},
  \bibinfo {author} {\bibfnamefont {J.}~\bibnamefont {Zeldes}}, \bibinfo
  {author} {\bibfnamefont {T.}~\bibnamefont {Enoto}}, \bibinfo {author}
  {\bibfnamefont {T.}~\bibnamefont {Okajima}}, \bibinfo {author} {\bibfnamefont
  {J.~W.}\ \bibnamefont {Kellogg}}, \bibinfo {author} {\bibfnamefont
  {C.}~\bibnamefont {Baker}}, \bibinfo {author} {\bibfnamefont {C.~B.}\
  \bibnamefont {Markwardt}}, \bibinfo {author} {\bibfnamefont {Z.}~\bibnamefont
  {Arzoumanian}}, \ and\ \bibinfo {author} {\bibfnamefont {K.~C.}\ \bibnamefont
  {Gendreau}},\ }\href {\doibase 10.3847/2041-8213/ab53eb} {\bibfield
  {journal} {\bibinfo  {journal} {The Astrophysical Journal}\ }\textbf
  {\bibinfo {volume} {887}},\ \bibinfo {pages} {L25} (\bibinfo {year}
  {2019}{\natexlab{a}})}\BibitemShut {NoStop}%
\bibitem [{\citenamefont {Bogdanov}\ \emph
  {et~al.}(2019{\natexlab{b}})\citenamefont {Bogdanov}, \citenamefont {Lamb},
  \citenamefont {Mahmoodifar}, \citenamefont {Miller}, \citenamefont {Morsink},
  \citenamefont {Riley}, \citenamefont {Strohmayer}, \citenamefont {Tung},
  \citenamefont {Watts}, \citenamefont {Dittmann}, \citenamefont {Chakrabarty},
  \citenamefont {Guillot}, \citenamefont {Arzoumanian},\ and\ \citenamefont
  {Gendreau}}]{Bogdanov_2019b}%
  \BibitemOpen
  \bibfield  {author} {\bibinfo {author} {\bibfnamefont {S.}~\bibnamefont
  {Bogdanov}}, \bibinfo {author} {\bibfnamefont {F.~K.}\ \bibnamefont {Lamb}},
  \bibinfo {author} {\bibfnamefont {S.}~\bibnamefont {Mahmoodifar}}, \bibinfo
  {author} {\bibfnamefont {M.~C.}\ \bibnamefont {Miller}}, \bibinfo {author}
  {\bibfnamefont {S.~M.}\ \bibnamefont {Morsink}}, \bibinfo {author}
  {\bibfnamefont {T.~E.}\ \bibnamefont {Riley}}, \bibinfo {author}
  {\bibfnamefont {T.~E.}\ \bibnamefont {Strohmayer}}, \bibinfo {author}
  {\bibfnamefont {A.~K.}\ \bibnamefont {Tung}}, \bibinfo {author}
  {\bibfnamefont {A.~L.}\ \bibnamefont {Watts}}, \bibinfo {author}
  {\bibfnamefont {A.~J.}\ \bibnamefont {Dittmann}}, \bibinfo {author}
  {\bibfnamefont {D.}~\bibnamefont {Chakrabarty}}, \bibinfo {author}
  {\bibfnamefont {S.}~\bibnamefont {Guillot}}, \bibinfo {author} {\bibfnamefont
  {Z.}~\bibnamefont {Arzoumanian}}, \ and\ \bibinfo {author} {\bibfnamefont
  {K.~C.}\ \bibnamefont {Gendreau}},\ }\href {\doibase
  10.3847/2041-8213/ab5968} {\bibfield  {journal} {\bibinfo  {journal} {The
  Astrophysical Journal}\ }\textbf {\bibinfo {volume} {887}},\ \bibinfo {pages}
  {L26} (\bibinfo {year} {2019}{\natexlab{b}})}\BibitemShut {NoStop}%
\bibitem [{\citenamefont {Guillot}\ \emph {et~al.}(2019)\citenamefont
  {Guillot}, \citenamefont {Kerr}, \citenamefont {Ray}, \citenamefont
  {Bogdanov}, \citenamefont {Ransom}, \citenamefont {Deneva}, \citenamefont
  {Arzoumanian}, \citenamefont {Bult}, \citenamefont {Chakrabarty},
  \citenamefont {Gendreau}, \citenamefont {Ho}, \citenamefont {Jaisawal},
  \citenamefont {Malacaria}, \citenamefont {Miller}, \citenamefont
  {Strohmayer}, \citenamefont {Wolff}, \citenamefont {Wood}, \citenamefont
  {Webb}, \citenamefont {Guillemot}, \citenamefont {Cognard},\ and\
  \citenamefont {Theureau}}]{Guillot_2019}%
  \BibitemOpen
  \bibfield  {author} {\bibinfo {author} {\bibfnamefont {S.}~\bibnamefont
  {Guillot}}, \bibinfo {author} {\bibfnamefont {M.}~\bibnamefont {Kerr}},
  \bibinfo {author} {\bibfnamefont {P.~S.}\ \bibnamefont {Ray}}, \bibinfo
  {author} {\bibfnamefont {S.}~\bibnamefont {Bogdanov}}, \bibinfo {author}
  {\bibfnamefont {S.}~\bibnamefont {Ransom}}, \bibinfo {author} {\bibfnamefont
  {J.~S.}\ \bibnamefont {Deneva}}, \bibinfo {author} {\bibfnamefont
  {Z.}~\bibnamefont {Arzoumanian}}, \bibinfo {author} {\bibfnamefont
  {P.}~\bibnamefont {Bult}}, \bibinfo {author} {\bibfnamefont {D.}~\bibnamefont
  {Chakrabarty}}, \bibinfo {author} {\bibfnamefont {K.~C.}\ \bibnamefont
  {Gendreau}}, \bibinfo {author} {\bibfnamefont {W.~C.~G.}\ \bibnamefont {Ho}},
  \bibinfo {author} {\bibfnamefont {G.~K.}\ \bibnamefont {Jaisawal}}, \bibinfo
  {author} {\bibfnamefont {C.}~\bibnamefont {Malacaria}}, \bibinfo {author}
  {\bibfnamefont {M.~C.}\ \bibnamefont {Miller}}, \bibinfo {author}
  {\bibfnamefont {T.~E.}\ \bibnamefont {Strohmayer}}, \bibinfo {author}
  {\bibfnamefont {M.~T.}\ \bibnamefont {Wolff}}, \bibinfo {author}
  {\bibfnamefont {K.~S.}\ \bibnamefont {Wood}}, \bibinfo {author}
  {\bibfnamefont {N.~A.}\ \bibnamefont {Webb}}, \bibinfo {author}
  {\bibfnamefont {L.}~\bibnamefont {Guillemot}}, \bibinfo {author}
  {\bibfnamefont {I.}~\bibnamefont {Cognard}}, \ and\ \bibinfo {author}
  {\bibfnamefont {G.}~\bibnamefont {Theureau}},\ }\href {\doibase
  10.3847/2041-8213/ab511b} {\bibfield  {journal} {\bibinfo  {journal} {The
  Astrophysical Journal}\ }\textbf {\bibinfo {volume} {887}},\ \bibinfo {pages}
  {L27} (\bibinfo {year} {2019})}\BibitemShut {NoStop}%
\bibitem [{\citenamefont {Raaijmakers}\ \emph
  {et~al.}(2019{\natexlab{a}})\citenamefont {Raaijmakers}, \citenamefont
  {Riley}, \citenamefont {Watts}, \citenamefont {Greif}, \citenamefont
  {Morsink}, \citenamefont {Hebeler}, \citenamefont {Schwenk}, \citenamefont
  {Hinderer}, \citenamefont {Nissanke}, \citenamefont {Guillot}, \citenamefont
  {Arzoumanian}, \citenamefont {Bogdanov}, \citenamefont {Chakrabarty},
  \citenamefont {Gendreau}, \citenamefont {Ho}, \citenamefont {Lattimer},
  \citenamefont {Ludlam},\ and\ \citenamefont {Wolff}}]{Raaijmakers_2019}%
  \BibitemOpen
  \bibfield  {author} {\bibinfo {author} {\bibfnamefont {G.}~\bibnamefont
  {Raaijmakers}}, \bibinfo {author} {\bibfnamefont {T.~E.}\ \bibnamefont
  {Riley}}, \bibinfo {author} {\bibfnamefont {A.~L.}\ \bibnamefont {Watts}},
  \bibinfo {author} {\bibfnamefont {S.~K.}\ \bibnamefont {Greif}}, \bibinfo
  {author} {\bibfnamefont {S.~M.}\ \bibnamefont {Morsink}}, \bibinfo {author}
  {\bibfnamefont {K.}~\bibnamefont {Hebeler}}, \bibinfo {author} {\bibfnamefont
  {A.}~\bibnamefont {Schwenk}}, \bibinfo {author} {\bibfnamefont
  {T.}~\bibnamefont {Hinderer}}, \bibinfo {author} {\bibfnamefont
  {S.}~\bibnamefont {Nissanke}}, \bibinfo {author} {\bibfnamefont
  {S.}~\bibnamefont {Guillot}}, \bibinfo {author} {\bibfnamefont
  {Z.}~\bibnamefont {Arzoumanian}}, \bibinfo {author} {\bibfnamefont
  {S.}~\bibnamefont {Bogdanov}}, \bibinfo {author} {\bibfnamefont
  {D.}~\bibnamefont {Chakrabarty}}, \bibinfo {author} {\bibfnamefont {K.~C.}\
  \bibnamefont {Gendreau}}, \bibinfo {author} {\bibfnamefont {W.~C.~G.}\
  \bibnamefont {Ho}}, \bibinfo {author} {\bibfnamefont {J.~M.}\ \bibnamefont
  {Lattimer}}, \bibinfo {author} {\bibfnamefont {R.~M.}\ \bibnamefont
  {Ludlam}}, \ and\ \bibinfo {author} {\bibfnamefont {M.~T.}\ \bibnamefont
  {Wolff}},\ }\href {\doibase 10.3847/2041-8213/ab451a} {\bibfield  {journal}
  {\bibinfo  {journal} {The Astrophysical Journal}\ }\textbf {\bibinfo {volume}
  {887}},\ \bibinfo {pages} {L22} (\bibinfo {year}
  {2019}{\natexlab{a}})}\BibitemShut {NoStop}%
\bibitem [{\citenamefont {Christian}\ and\ \citenamefont
  {Schaffner-Bielich}(2019)}]{Christian:2019qer}%
  \BibitemOpen
  \bibfield  {author} {\bibinfo {author} {\bibfnamefont {J.-E.}\ \bibnamefont
  {Christian}}\ and\ \bibinfo {author} {\bibfnamefont {J.}~\bibnamefont
  {Schaffner-Bielich}},\ }\href@noop {} {\  (\bibinfo {year} {2019})},\ \Eprint
  {http://arxiv.org/abs/1912.09809} {arXiv:1912.09809 [astro-ph.HE]}
  \BibitemShut {NoStop}%
\bibitem [{\citenamefont {Jiang}\ \emph {et~al.}(2019)\citenamefont {Jiang},
  \citenamefont {Tang}, \citenamefont {Wang}, \citenamefont {Fan},\ and\
  \citenamefont {Wei}}]{Jiang:2019rcw}%
  \BibitemOpen
  \bibfield  {author} {\bibinfo {author} {\bibfnamefont {J.-L.}\ \bibnamefont
  {Jiang}}, \bibinfo {author} {\bibfnamefont {S.-P.}\ \bibnamefont {Tang}},
  \bibinfo {author} {\bibfnamefont {Y.-Z.}\ \bibnamefont {Wang}}, \bibinfo
  {author} {\bibfnamefont {Y.-Z.}\ \bibnamefont {Fan}}, \ and\ \bibinfo
  {author} {\bibfnamefont {D.-M.}\ \bibnamefont {Wei}},\ }\href@noop {} {\
  (\bibinfo {year} {2019})},\ \Eprint {http://arxiv.org/abs/1912.07467}
  {arXiv:1912.07467 [astro-ph.HE]} \BibitemShut {NoStop}%
\bibitem [{\citenamefont {Raaijmakers}\ \emph
  {et~al.}(2019{\natexlab{b}})\citenamefont {Raaijmakers} \emph
  {et~al.}}]{Raaijmakers:2019dks}%
  \BibitemOpen
  \bibfield  {author} {\bibinfo {author} {\bibfnamefont {G.}~\bibnamefont
  {Raaijmakers}} \emph {et~al.},\ }\href@noop {} {\  (\bibinfo {year}
  {2019}{\natexlab{b}})},\ \Eprint {http://arxiv.org/abs/1912.11031}
  {arXiv:1912.11031 [astro-ph.HE]} \BibitemShut {NoStop}%
\bibitem [{\citenamefont {Zimmerman}\ \emph {et~al.}(2020)\citenamefont
  {Zimmerman}, \citenamefont {Carson}, \citenamefont {Schumacher},
  \citenamefont {Steiner},\ and\ \citenamefont {Yagi}}]{Zimmerman:2020eho}%
  \BibitemOpen
  \bibfield  {author} {\bibinfo {author} {\bibfnamefont {J.}~\bibnamefont
  {Zimmerman}}, \bibinfo {author} {\bibfnamefont {Z.}~\bibnamefont {Carson}},
  \bibinfo {author} {\bibfnamefont {K.}~\bibnamefont {Schumacher}}, \bibinfo
  {author} {\bibfnamefont {A.~W.}\ \bibnamefont {Steiner}}, \ and\ \bibinfo
  {author} {\bibfnamefont {K.}~\bibnamefont {Yagi}},\ }\href@noop {} {\
  (\bibinfo {year} {2020})},\ \Eprint {http://arxiv.org/abs/2002.03210}
  {arXiv:2002.03210 [astro-ph.HE]} \BibitemShut {NoStop}%
\bibitem [{\citenamefont {Dietrich}\ \emph {et~al.}(2020)\citenamefont
  {Dietrich}, \citenamefont {Coughlin}, \citenamefont {Pang}, \citenamefont
  {Bulla}, \citenamefont {Heinzel}, \citenamefont {Issa}, \citenamefont
  {Tews},\ and\ \citenamefont {Antier}}]{Dietrich:2020lps}%
  \BibitemOpen
  \bibfield  {author} {\bibinfo {author} {\bibfnamefont {T.}~\bibnamefont
  {Dietrich}}, \bibinfo {author} {\bibfnamefont {M.~W.}\ \bibnamefont
  {Coughlin}}, \bibinfo {author} {\bibfnamefont {P.~T.~H.}\ \bibnamefont
  {Pang}}, \bibinfo {author} {\bibfnamefont {M.}~\bibnamefont {Bulla}},
  \bibinfo {author} {\bibfnamefont {J.}~\bibnamefont {Heinzel}}, \bibinfo
  {author} {\bibfnamefont {L.}~\bibnamefont {Issa}}, \bibinfo {author}
  {\bibfnamefont {I.}~\bibnamefont {Tews}}, \ and\ \bibinfo {author}
  {\bibfnamefont {S.}~\bibnamefont {Antier}},\ }\href@noop {} {\  (\bibinfo
  {year} {2020})},\ \Eprint {http://arxiv.org/abs/2002.11355} {arXiv:2002.11355
  [astro-ph.HE]} \BibitemShut {NoStop}%
\bibitem [{\citenamefont {{Demorest}}\ \emph {et~al.}(2010)\citenamefont
  {{Demorest}}, \citenamefont {{Pennucci}}, \citenamefont {{Ransom}},
  \citenamefont {{Roberts}},\ and\ \citenamefont {{Hessels}}}]{1.97NS}%
  \BibitemOpen
  \bibfield  {author} {\bibinfo {author} {\bibfnamefont {P.~B.}\ \bibnamefont
  {{Demorest}}}, \bibinfo {author} {\bibfnamefont {T.}~\bibnamefont
  {{Pennucci}}}, \bibinfo {author} {\bibfnamefont {S.~M.}\ \bibnamefont
  {{Ransom}}}, \bibinfo {author} {\bibfnamefont {M.~S.~E.}\ \bibnamefont
  {{Roberts}}}, \ and\ \bibinfo {author} {\bibfnamefont {J.~W.~T.}\
  \bibnamefont {{Hessels}}},\ }\href {\doibase 10.1038/nature09466} {\bibfield
  {journal} {\bibinfo  {journal} {Nature}\ }\textbf {\bibinfo {volume} {467}},\
  \bibinfo {pages} {1081} (\bibinfo {year} {2010})},\ \Eprint
  {http://arxiv.org/abs/1010.5788} {arXiv:1010.5788 [astro-ph.HE]} \BibitemShut
  {NoStop}%
\bibitem [{\citenamefont {Antoniadis}\ \emph {et~al.}(2013)\citenamefont
  {Antoniadis}, \citenamefont {Freire}, \citenamefont {Wex}, \citenamefont
  {Tauris}, \citenamefont {Lynch} \emph {et~al.}}]{2.01NS}%
  \BibitemOpen
  \bibfield  {author} {\bibinfo {author} {\bibfnamefont {J.}~\bibnamefont
  {Antoniadis}}, \bibinfo {author} {\bibfnamefont {P.~C.}\ \bibnamefont
  {Freire}}, \bibinfo {author} {\bibfnamefont {N.}~\bibnamefont {Wex}},
  \bibinfo {author} {\bibfnamefont {T.~M.}\ \bibnamefont {Tauris}}, \bibinfo
  {author} {\bibfnamefont {R.~S.}\ \bibnamefont {Lynch}},  \emph {et~al.},\
  }\href {\doibase 10.1126/science.1233232} {\bibfield  {journal} {\bibinfo
  {journal} {Science}\ }\textbf {\bibinfo {volume} {340}},\ \bibinfo {pages}
  {1233232} (\bibinfo {year} {2013})},\ \Eprint
  {http://arxiv.org/abs/1304.6875} {arXiv:1304.6875 [astro-ph.HE]} \BibitemShut
  {NoStop}%
\bibitem [{\citenamefont {Cromartie}\ \emph {et~al.}(2020)\citenamefont
  {Cromartie}, \citenamefont {Fonseca}, \citenamefont {Ransom}, \citenamefont
  {Demorest}, \citenamefont {Arzoumanian}, \citenamefont {Blumer},
  \citenamefont {Brook}, \citenamefont {DeCesar}, \citenamefont {Dolch},
  \citenamefont {Ellis}, \citenamefont {Ferdman}, \citenamefont {Ferrara},
  \citenamefont {Garver-Daniels}, \citenamefont {Gentile}, \citenamefont
  {Jones}, \citenamefont {Lam}, \citenamefont {Lorimer}, \citenamefont {Lynch},
  \citenamefont {McLaughlin}, \citenamefont {Ng}, \citenamefont {Nice},
  \citenamefont {Pennucci}, \citenamefont {Spiewak}, \citenamefont {Stairs},
  \citenamefont {Stovall}, \citenamefont {Swiggum},\ and\ \citenamefont
  {Zhu}}]{Thankful}%
  \BibitemOpen
  \bibfield  {author} {\bibinfo {author} {\bibfnamefont {H.~T.}\ \bibnamefont
  {Cromartie}}, \bibinfo {author} {\bibfnamefont {E.}~\bibnamefont {Fonseca}},
  \bibinfo {author} {\bibfnamefont {S.~M.}\ \bibnamefont {Ransom}}, \bibinfo
  {author} {\bibfnamefont {P.~B.}\ \bibnamefont {Demorest}}, \bibinfo {author}
  {\bibfnamefont {Z.}~\bibnamefont {Arzoumanian}}, \bibinfo {author}
  {\bibfnamefont {H.}~\bibnamefont {Blumer}}, \bibinfo {author} {\bibfnamefont
  {P.~R.}\ \bibnamefont {Brook}}, \bibinfo {author} {\bibfnamefont {M.~E.}\
  \bibnamefont {DeCesar}}, \bibinfo {author} {\bibfnamefont {T.}~\bibnamefont
  {Dolch}}, \bibinfo {author} {\bibfnamefont {J.~A.}\ \bibnamefont {Ellis}},
  \bibinfo {author} {\bibfnamefont {R.~D.}\ \bibnamefont {Ferdman}}, \bibinfo
  {author} {\bibfnamefont {E.~C.}\ \bibnamefont {Ferrara}}, \bibinfo {author}
  {\bibfnamefont {N.}~\bibnamefont {Garver-Daniels}}, \bibinfo {author}
  {\bibfnamefont {P.~A.}\ \bibnamefont {Gentile}}, \bibinfo {author}
  {\bibfnamefont {M.~L.}\ \bibnamefont {Jones}}, \bibinfo {author}
  {\bibfnamefont {M.~T.}\ \bibnamefont {Lam}}, \bibinfo {author} {\bibfnamefont
  {D.~R.}\ \bibnamefont {Lorimer}}, \bibinfo {author} {\bibfnamefont {R.~S.}\
  \bibnamefont {Lynch}}, \bibinfo {author} {\bibfnamefont {M.~A.}\ \bibnamefont
  {McLaughlin}}, \bibinfo {author} {\bibfnamefont {C.}~\bibnamefont {Ng}},
  \bibinfo {author} {\bibfnamefont {D.~J.}\ \bibnamefont {Nice}}, \bibinfo
  {author} {\bibfnamefont {T.~T.}\ \bibnamefont {Pennucci}}, \bibinfo {author}
  {\bibfnamefont {R.}~\bibnamefont {Spiewak}}, \bibinfo {author} {\bibfnamefont
  {I.~H.}\ \bibnamefont {Stairs}}, \bibinfo {author} {\bibfnamefont
  {K.}~\bibnamefont {Stovall}}, \bibinfo {author} {\bibfnamefont {J.~K.}\
  \bibnamefont {Swiggum}}, \ and\ \bibinfo {author} {\bibfnamefont {W.~W.}\
  \bibnamefont {Zhu}},\ }\href {\doibase 10.1038/s41550-019-0880-2} {\bibfield
  {journal} {\bibinfo  {journal} {Nature Astronomy}\ }\textbf {\bibinfo
  {volume} {4}},\ \bibinfo {pages} {72} (\bibinfo {year} {2020})}\BibitemShut
  {NoStop}%
\bibitem [{\citenamefont {Kramer}\ and\ \citenamefont
  {Wex}(2009)}]{Kramer_2009}%
  \BibitemOpen
  \bibfield  {author} {\bibinfo {author} {\bibfnamefont {M.}~\bibnamefont
  {Kramer}}\ and\ \bibinfo {author} {\bibfnamefont {N.}~\bibnamefont {Wex}},\
  }\href {\doibase 10.1088/0264-9381/26/7/073001} {\bibfield  {journal}
  {\bibinfo  {journal} {Classical and Quantum Gravity}\ }\textbf {\bibinfo
  {volume} {26}},\ \bibinfo {pages} {073001} (\bibinfo {year}
  {2009})}\BibitemShut {NoStop}%
\bibitem [{\citenamefont {Lattimer}\ and\ \citenamefont
  {Schutz}(2005)}]{Lattimer_2005}%
  \BibitemOpen
  \bibfield  {author} {\bibinfo {author} {\bibfnamefont {J.~M.}\ \bibnamefont
  {Lattimer}}\ and\ \bibinfo {author} {\bibfnamefont {B.~F.}\ \bibnamefont
  {Schutz}},\ }\href {\doibase 10.1086/431543} {\bibfield  {journal} {\bibinfo
  {journal} {The Astrophysical Journal}\ }\textbf {\bibinfo {volume} {629}},\
  \bibinfo {pages} {979} (\bibinfo {year} {2005})}\BibitemShut {NoStop}%
\bibitem [{\citenamefont {Hinderer}(2008)}]{Hinderer_2008}%
  \BibitemOpen
  \bibfield  {author} {\bibinfo {author} {\bibfnamefont {T.}~\bibnamefont
  {Hinderer}},\ }\href {\doibase 10.1086/533487} {\bibfield  {journal}
  {\bibinfo  {journal} {The Astrophysical Journal}\ }\textbf {\bibinfo {volume}
  {677}},\ \bibinfo {pages} {1216} (\bibinfo {year} {2008})}\BibitemShut
  {NoStop}%
\bibitem [{\citenamefont {Flanagan}\ and\ \citenamefont
  {Hinderer}(2008)}]{PhysRevD.77.021502}%
  \BibitemOpen
  \bibfield  {author} {\bibinfo {author} {\bibfnamefont {E.~E.}\ \bibnamefont
  {Flanagan}}\ and\ \bibinfo {author} {\bibfnamefont {T.}~\bibnamefont
  {Hinderer}},\ }\href {\doibase 10.1103/PhysRevD.77.021502} {\bibfield
  {journal} {\bibinfo  {journal} {Phys. Rev. D}\ }\textbf {\bibinfo {volume}
  {77}},\ \bibinfo {pages} {021502} (\bibinfo {year} {2008})}\BibitemShut
  {NoStop}%
\bibitem [{\citenamefont {Abbott}\ \emph
  {et~al.}(2018{\natexlab{a}})\citenamefont {Abbott}, \citenamefont {Abbott},
  \citenamefont {Abbott}, \citenamefont {Acernese}, \citenamefont {Ackley},
  \citenamefont {Adams}, \citenamefont {Adams}, \citenamefont {Addesso},
  \citenamefont {Adhikari}, \citenamefont {Adya},\ and\ \citenamefont
  {et~al.}}]{Abbott_2018}%
  \BibitemOpen
  \bibfield  {author} {\bibinfo {author} {\bibfnamefont {B.}~\bibnamefont
  {Abbott}}, \bibinfo {author} {\bibfnamefont {R.}~\bibnamefont {Abbott}},
  \bibinfo {author} {\bibfnamefont {T.}~\bibnamefont {Abbott}}, \bibinfo
  {author} {\bibfnamefont {F.}~\bibnamefont {Acernese}}, \bibinfo {author}
  {\bibfnamefont {K.}~\bibnamefont {Ackley}}, \bibinfo {author} {\bibfnamefont
  {C.}~\bibnamefont {Adams}}, \bibinfo {author} {\bibfnamefont
  {T.}~\bibnamefont {Adams}}, \bibinfo {author} {\bibfnamefont
  {P.}~\bibnamefont {Addesso}}, \bibinfo {author} {\bibfnamefont
  {R.}~\bibnamefont {Adhikari}}, \bibinfo {author} {\bibfnamefont
  {V.}~\bibnamefont {Adya}}, \ and\ \bibinfo {author} {\bibnamefont {et~al.}},\
  }\href {\doibase 10.1103/physrevlett.121.161101} {\bibfield  {journal}
  {\bibinfo  {journal} {Physical Review Letters}\ }\textbf {\bibinfo {volume}
  {121}} (\bibinfo {year} {2018}{\natexlab{a}}),\
  10.1103/physrevlett.121.161101}\BibitemShut {NoStop}%
\bibitem [{\citenamefont {Abbott}\ \emph
  {et~al.}(2018{\natexlab{b}})\citenamefont {Abbott} \emph
  {et~al.}}]{LIGO:posterior}%
  \BibitemOpen
  \bibfield  {author} {\bibinfo {author} {\bibfnamefont {B.~P.}\ \bibnamefont
  {Abbott}} \emph {et~al.} (\bibinfo {collaboration} {LIGO Scientific,
  Virgo}),\ }\href {\doibase 10.1103/PhysRevLett.121.161101} {\bibfield
  {journal} {\bibinfo  {journal} {Phys. Rev. Lett.}\ }\textbf {\bibinfo
  {volume} {121}},\ \bibinfo {pages} {161101} (\bibinfo {year}
  {2018}{\natexlab{b}})},\ \Eprint {http://arxiv.org/abs/1805.11581}
  {arXiv:1805.11581 [gr-qc]} \BibitemShut {NoStop}%
\bibitem [{\citenamefont {Annala}\ \emph {et~al.}(2018)\citenamefont {Annala},
  \citenamefont {Gorda}, \citenamefont {Kurkela},\ and\ \citenamefont
  {Vuorinen}}]{Annala:2017llu}%
  \BibitemOpen
  \bibfield  {author} {\bibinfo {author} {\bibfnamefont {E.}~\bibnamefont
  {Annala}}, \bibinfo {author} {\bibfnamefont {T.}~\bibnamefont {Gorda}},
  \bibinfo {author} {\bibfnamefont {A.}~\bibnamefont {Kurkela}}, \ and\
  \bibinfo {author} {\bibfnamefont {A.}~\bibnamefont {Vuorinen}},\ }\href
  {\doibase 10.1103/PhysRevLett.120.172703} {\bibfield  {journal} {\bibinfo
  {journal} {Phys. Rev. Lett.}\ }\textbf {\bibinfo {volume} {120}},\ \bibinfo
  {pages} {172703} (\bibinfo {year} {2018})},\ \Eprint
  {http://arxiv.org/abs/1711.02644} {arXiv:1711.02644 [astro-ph.HE]}
  \BibitemShut {NoStop}%
\bibitem [{\citenamefont {Abbott}\ \emph
  {et~al.}(2018{\natexlab{c}})\citenamefont {Abbott} \emph
  {et~al.}}]{Abbott:2018exr}%
  \BibitemOpen
  \bibfield  {author} {\bibinfo {author} {\bibfnamefont {B.~P.}\ \bibnamefont
  {Abbott}} \emph {et~al.} (\bibinfo {collaboration} {LIGO Scientific,
  Virgo}),\ }\href {\doibase 10.1103/PhysRevLett.121.161101} {\bibfield
  {journal} {\bibinfo  {journal} {Phys. Rev. Lett.}\ }\textbf {\bibinfo
  {volume} {121}},\ \bibinfo {pages} {161101} (\bibinfo {year}
  {2018}{\natexlab{c}})},\ \Eprint {http://arxiv.org/abs/1805.11581}
  {arXiv:1805.11581 [gr-qc]} \BibitemShut {NoStop}%
\bibitem [{\citenamefont {Raithel}\ \emph {et~al.}(2018)\citenamefont
  {Raithel}, \citenamefont {Ozel},\ and\ \citenamefont
  {Psaltis}}]{Raithel:2018ncd}%
  \BibitemOpen
  \bibfield  {author} {\bibinfo {author} {\bibfnamefont {C.}~\bibnamefont
  {Raithel}}, \bibinfo {author} {\bibfnamefont {F.}~\bibnamefont {Ozel}}, \
  and\ \bibinfo {author} {\bibfnamefont {D.}~\bibnamefont {Psaltis}},\ }\href
  {\doibase 10.3847/2041-8213/aabcbf} {\bibfield  {journal} {\bibinfo
  {journal} {Astrophys. J.}\ }\textbf {\bibinfo {volume} {857}},\ \bibinfo
  {pages} {L23} (\bibinfo {year} {2018})},\ \Eprint
  {http://arxiv.org/abs/1803.07687} {arXiv:1803.07687 [astro-ph.HE]}
  \BibitemShut {NoStop}%
\bibitem [{\citenamefont {Lim}\ and\ \citenamefont {Holt}(2018)}]{Lim:2018bkq}%
  \BibitemOpen
  \bibfield  {author} {\bibinfo {author} {\bibfnamefont {Y.}~\bibnamefont
  {Lim}}\ and\ \bibinfo {author} {\bibfnamefont {J.~W.}\ \bibnamefont {Holt}},\
  }\href {\doibase 10.1103/PhysRevLett.121.062701} {\bibfield  {journal}
  {\bibinfo  {journal} {Phys. Rev. Lett.}\ }\textbf {\bibinfo {volume} {121}},\
  \bibinfo {pages} {062701} (\bibinfo {year} {2018})},\ \Eprint
  {http://arxiv.org/abs/1803.02803} {arXiv:1803.02803 [nucl-th]} \BibitemShut
  {NoStop}%
\bibitem [{\citenamefont {Bauswein}\ \emph {et~al.}(2017)\citenamefont
  {Bauswein}, \citenamefont {Just}, \citenamefont {Janka},\ and\ \citenamefont
  {Stergioulas}}]{Bauswein:2017vtn}%
  \BibitemOpen
  \bibfield  {author} {\bibinfo {author} {\bibfnamefont {A.}~\bibnamefont
  {Bauswein}}, \bibinfo {author} {\bibfnamefont {O.}~\bibnamefont {Just}},
  \bibinfo {author} {\bibfnamefont {H.-T.}\ \bibnamefont {Janka}}, \ and\
  \bibinfo {author} {\bibfnamefont {N.}~\bibnamefont {Stergioulas}},\ }\href
  {\doibase 10.3847/2041-8213/aa9994} {\bibfield  {journal} {\bibinfo
  {journal} {Astrophys. J.}\ }\textbf {\bibinfo {volume} {850}},\ \bibinfo
  {pages} {L34} (\bibinfo {year} {2017})},\ \Eprint
  {http://arxiv.org/abs/1710.06843} {arXiv:1710.06843 [astro-ph.HE]}
  \BibitemShut {NoStop}%
\bibitem [{\citenamefont {De}\ \emph {et~al.}(2018)\citenamefont {De},
  \citenamefont {Finstad}, \citenamefont {Lattimer}, \citenamefont {Brown},
  \citenamefont {Berger},\ and\ \citenamefont {Biwer}}]{De:2018uhw}%
  \BibitemOpen
  \bibfield  {author} {\bibinfo {author} {\bibfnamefont {S.}~\bibnamefont
  {De}}, \bibinfo {author} {\bibfnamefont {D.}~\bibnamefont {Finstad}},
  \bibinfo {author} {\bibfnamefont {J.~M.}\ \bibnamefont {Lattimer}}, \bibinfo
  {author} {\bibfnamefont {D.~A.}\ \bibnamefont {Brown}}, \bibinfo {author}
  {\bibfnamefont {E.}~\bibnamefont {Berger}}, \ and\ \bibinfo {author}
  {\bibfnamefont {C.~M.}\ \bibnamefont {Biwer}},\ }\href {\doibase
  10.1103/PhysRevLett.121.259902, 10.1103/PhysRevLett.121.091102} {\bibfield
  {journal} {\bibinfo  {journal} {Phys. Rev. Lett.}\ }\textbf {\bibinfo
  {volume} {121}},\ \bibinfo {pages} {091102} (\bibinfo {year} {2018})},\
  \bibinfo {note} {[Erratum: Phys. Rev. Lett.121,no.25,259902(2018)]},\ \Eprint
  {http://arxiv.org/abs/1804.08583} {arXiv:1804.08583 [astro-ph.HE]}
  \BibitemShut {NoStop}%
\bibitem [{\citenamefont {Most}\ \emph {et~al.}(2018)\citenamefont {Most},
  \citenamefont {Weih}, \citenamefont {Rezzolla},\ and\ \citenamefont
  {Schaffner-Bielich}}]{Most:2018hfd}%
  \BibitemOpen
  \bibfield  {author} {\bibinfo {author} {\bibfnamefont {E.~R.}\ \bibnamefont
  {Most}}, \bibinfo {author} {\bibfnamefont {L.~R.}\ \bibnamefont {Weih}},
  \bibinfo {author} {\bibfnamefont {L.}~\bibnamefont {Rezzolla}}, \ and\
  \bibinfo {author} {\bibfnamefont {J.}~\bibnamefont {Schaffner-Bielich}},\
  }\href {\doibase 10.1103/PhysRevLett.120.261103} {\bibfield  {journal}
  {\bibinfo  {journal} {Phys. Rev. Lett.}\ }\textbf {\bibinfo {volume} {120}},\
  \bibinfo {pages} {261103} (\bibinfo {year} {2018})},\ \Eprint
  {http://arxiv.org/abs/1803.00549} {arXiv:1803.00549 [gr-qc]} \BibitemShut
  {NoStop}%
\bibitem [{\citenamefont {Annala}\ \emph {et~al.}(2019)\citenamefont {Annala},
  \citenamefont {Gorda}, \citenamefont {Kurkela}, \citenamefont {Nättilä},\
  and\ \citenamefont {Vuorinen}}]{Annala:2019puf}%
  \BibitemOpen
  \bibfield  {author} {\bibinfo {author} {\bibfnamefont {E.}~\bibnamefont
  {Annala}}, \bibinfo {author} {\bibfnamefont {T.}~\bibnamefont {Gorda}},
  \bibinfo {author} {\bibfnamefont {A.}~\bibnamefont {Kurkela}}, \bibinfo
  {author} {\bibfnamefont {J.}~\bibnamefont {Nättilä}}, \ and\ \bibinfo
  {author} {\bibfnamefont {A.}~\bibnamefont {Vuorinen}},\ }\href@noop {} {\
  (\bibinfo {year} {2019})},\ \Eprint {http://arxiv.org/abs/1903.09121}
  {arXiv:1903.09121 [astro-ph.HE]} \BibitemShut {NoStop}%
\bibitem [{\citenamefont {Malik}\ \emph {et~al.}(2018)\citenamefont {Malik},
  \citenamefont {Alam}, \citenamefont {Fortin}, \citenamefont {Providência},
  \citenamefont {Agrawal}, \citenamefont {Jha}, \citenamefont {Kumar},\ and\
  \citenamefont {Patra}}]{Malik2018}%
  \BibitemOpen
  \bibfield  {author} {\bibinfo {author} {\bibfnamefont {T.}~\bibnamefont
  {Malik}}, \bibinfo {author} {\bibfnamefont {N.}~\bibnamefont {Alam}},
  \bibinfo {author} {\bibfnamefont {M.}~\bibnamefont {Fortin}}, \bibinfo
  {author} {\bibfnamefont {C.}~\bibnamefont {Providência}}, \bibinfo {author}
  {\bibfnamefont {B.~K.}\ \bibnamefont {Agrawal}}, \bibinfo {author}
  {\bibfnamefont {T.~K.}\ \bibnamefont {Jha}}, \bibinfo {author} {\bibfnamefont
  {B.}~\bibnamefont {Kumar}}, \ and\ \bibinfo {author} {\bibfnamefont {S.~K.}\
  \bibnamefont {Patra}},\ }\href {\doibase 10.1103/PhysRevC.98.035804}
  {\bibfield  {journal} {\bibinfo  {journal} {Phys. Rev.}\ }\textbf {\bibinfo
  {volume} {C98}},\ \bibinfo {pages} {035804} (\bibinfo {year} {2018})},\
  \Eprint {http://arxiv.org/abs/1805.11963} {arXiv:1805.11963 [nucl-th]}
  \BibitemShut {NoStop}%
\bibitem [{\citenamefont {Carson}\ \emph
  {et~al.}(2019{\natexlab{a}})\citenamefont {Carson}, \citenamefont {Steiner},\
  and\ \citenamefont {Yagi}}]{Zack:nuclearConstraints}%
  \BibitemOpen
  \bibfield  {author} {\bibinfo {author} {\bibfnamefont {Z.}~\bibnamefont
  {Carson}}, \bibinfo {author} {\bibfnamefont {A.~W.}\ \bibnamefont {Steiner}},
  \ and\ \bibinfo {author} {\bibfnamefont {K.}~\bibnamefont {Yagi}},\ }\href
  {\doibase 10.1103/PhysRevD.99.043010} {\bibfield  {journal} {\bibinfo
  {journal} {Phys. Rev. D}\ }\textbf {\bibinfo {volume} {99}},\ \bibinfo
  {pages} {043010} (\bibinfo {year} {2019}{\natexlab{a}})}\BibitemShut
  {NoStop}%
\bibitem [{\citenamefont {Carson}\ \emph
  {et~al.}(2019{\natexlab{b}})\citenamefont {Carson}, \citenamefont {Steiner},\
  and\ \citenamefont {Yagi}}]{Carson:2019xxz}%
  \BibitemOpen
  \bibfield  {author} {\bibinfo {author} {\bibfnamefont {Z.}~\bibnamefont
  {Carson}}, \bibinfo {author} {\bibfnamefont {A.~W.}\ \bibnamefont {Steiner}},
  \ and\ \bibinfo {author} {\bibfnamefont {K.}~\bibnamefont {Yagi}},\ }\href
  {\doibase 10.1103/PhysRevD.100.023012} {\bibfield  {journal} {\bibinfo
  {journal} {Phys. Rev.}\ }\textbf {\bibinfo {volume} {D100}},\ \bibinfo
  {pages} {023012} (\bibinfo {year} {2019}{\natexlab{b}})},\ \Eprint
  {http://arxiv.org/abs/1906.05978} {arXiv:1906.05978 [gr-qc]} \BibitemShut
  {NoStop}%
\bibitem [{\citenamefont {Raithel}\ and\ \citenamefont
  {Ozel}(2019)}]{Raithel:2019ejc}%
  \BibitemOpen
  \bibfield  {author} {\bibinfo {author} {\bibfnamefont {C.~A.}\ \bibnamefont
  {Raithel}}\ and\ \bibinfo {author} {\bibfnamefont {F.}~\bibnamefont {Ozel}},\
  }\href {\doibase 10.3847/1538-4357/ab48e6} {\  (\bibinfo {year} {2019}),\
  10.3847/1538-4357/ab48e6},\ \Eprint {http://arxiv.org/abs/1908.00018}
  {arXiv:1908.00018 [astro-ph.HE]} \BibitemShut {NoStop}%
\bibitem [{\citenamefont {Yagi}\ and\ \citenamefont
  {Yunes}(2017{\natexlab{a}})}]{YAGI20171}%
  \BibitemOpen
  \bibfield  {author} {\bibinfo {author} {\bibfnamefont {K.}~\bibnamefont
  {Yagi}}\ and\ \bibinfo {author} {\bibfnamefont {N.}~\bibnamefont {Yunes}},\
  }\href {\doibase https://doi.org/10.1016/j.physrep.2017.03.002} {\bibfield
  {journal} {\bibinfo  {journal} {Physics Reports}\ }\textbf {\bibinfo {volume}
  {681}},\ \bibinfo {pages} {1 } (\bibinfo {year}
  {2017}{\natexlab{a}})}\BibitemShut {NoStop}%
\bibitem [{\citenamefont {Doneva}\ and\ \citenamefont
  {Pappas}(2018)}]{Doneva:2017jop}%
  \BibitemOpen
  \bibfield  {author} {\bibinfo {author} {\bibfnamefont {D.~D.}\ \bibnamefont
  {Doneva}}\ and\ \bibinfo {author} {\bibfnamefont {G.}~\bibnamefont
  {Pappas}},\ }\href {\doibase 10.1007/978-3-319-97616-7_13} {\bibfield
  {journal} {\bibinfo  {journal} {Astrophys. Space Sci. Libr.}\ }\textbf
  {\bibinfo {volume} {457}},\ \bibinfo {pages} {737} (\bibinfo {year}
  {2018})},\ \Eprint {http://arxiv.org/abs/1709.08046} {arXiv:1709.08046
  [gr-qc]} \BibitemShut {NoStop}%
\bibitem [{\citenamefont {Yagi}\ and\ \citenamefont
  {Yunes}(2013{\natexlab{a}})}]{Yagi365}%
  \BibitemOpen
  \bibfield  {author} {\bibinfo {author} {\bibfnamefont {K.}~\bibnamefont
  {Yagi}}\ and\ \bibinfo {author} {\bibfnamefont {N.}~\bibnamefont {Yunes}},\
  }\href {\doibase 10.1126/science.1236462} {\bibfield  {journal} {\bibinfo
  {journal} {Science}\ }\textbf {\bibinfo {volume} {341}},\ \bibinfo {pages}
  {365} (\bibinfo {year} {2013}{\natexlab{a}})}\BibitemShut {NoStop}%
\bibitem [{\citenamefont {Yagi}\ and\ \citenamefont
  {Yunes}(2013{\natexlab{b}})}]{Yagi:2013awa}%
  \BibitemOpen
  \bibfield  {author} {\bibinfo {author} {\bibfnamefont {K.}~\bibnamefont
  {Yagi}}\ and\ \bibinfo {author} {\bibfnamefont {N.}~\bibnamefont {Yunes}},\
  }\href {\doibase 10.1103/PhysRevD.88.023009} {\bibfield  {journal} {\bibinfo
  {journal} {Phys. Rev.}\ }\textbf {\bibinfo {volume} {D88}},\ \bibinfo {pages}
  {023009} (\bibinfo {year} {2013}{\natexlab{b}})},\ \Eprint
  {http://arxiv.org/abs/1303.1528} {arXiv:1303.1528 [gr-qc]} \BibitemShut
  {NoStop}%
\bibitem [{\citenamefont {Yagi}\ and\ \citenamefont {Yunes}(2016)}]{Yagi_2016}%
  \BibitemOpen
  \bibfield  {author} {\bibinfo {author} {\bibfnamefont {K.}~\bibnamefont
  {Yagi}}\ and\ \bibinfo {author} {\bibfnamefont {N.}~\bibnamefont {Yunes}},\
  }\href {\doibase 10.1088/0264-9381/33/13/13lt01} {\bibfield  {journal}
  {\bibinfo  {journal} {Classical and Quantum Gravity}\ }\textbf {\bibinfo
  {volume} {33}},\ \bibinfo {pages} {13LT01} (\bibinfo {year}
  {2016})}\BibitemShut {NoStop}%
\bibitem [{\citenamefont {Yagi}\ and\ \citenamefont
  {Yunes}(2017{\natexlab{b}})}]{Yagi:2016qmr}%
  \BibitemOpen
  \bibfield  {author} {\bibinfo {author} {\bibfnamefont {K.}~\bibnamefont
  {Yagi}}\ and\ \bibinfo {author} {\bibfnamefont {N.}~\bibnamefont {Yunes}},\
  }\href {\doibase 10.1088/1361-6382/34/1/015006} {\bibfield  {journal}
  {\bibinfo  {journal} {Class. Quant. Grav.}\ }\textbf {\bibinfo {volume}
  {34}},\ \bibinfo {pages} {015006} (\bibinfo {year} {2017}{\natexlab{b}})},\
  \Eprint {http://arxiv.org/abs/1608.06187} {arXiv:1608.06187 [gr-qc]}
  \BibitemShut {NoStop}%
\bibitem [{\citenamefont {Yagi}\ \emph
  {et~al.}(2014{\natexlab{a}})\citenamefont {Yagi}, \citenamefont {Stein},
  \citenamefont {Pappas}, \citenamefont {Yunes},\ and\ \citenamefont
  {Apostolatos}}]{Yagi:2014qua}%
  \BibitemOpen
  \bibfield  {author} {\bibinfo {author} {\bibfnamefont {K.}~\bibnamefont
  {Yagi}}, \bibinfo {author} {\bibfnamefont {L.~C.}\ \bibnamefont {Stein}},
  \bibinfo {author} {\bibfnamefont {G.}~\bibnamefont {Pappas}}, \bibinfo
  {author} {\bibfnamefont {N.}~\bibnamefont {Yunes}}, \ and\ \bibinfo {author}
  {\bibfnamefont {T.~A.}\ \bibnamefont {Apostolatos}},\ }\href {\doibase
  10.1103/PhysRevD.90.063010} {\bibfield  {journal} {\bibinfo  {journal} {Phys.
  Rev.}\ }\textbf {\bibinfo {volume} {D90}},\ \bibinfo {pages} {063010}
  (\bibinfo {year} {2014}{\natexlab{a}})},\ \Eprint
  {http://arxiv.org/abs/1406.7587} {arXiv:1406.7587 [gr-qc]} \BibitemShut
  {NoStop}%
\bibitem [{\citenamefont {Sham}\ \emph {et~al.}(2015)\citenamefont {Sham},
  \citenamefont {Chan}, \citenamefont {Lin},\ and\ \citenamefont
  {Leung}}]{Sham_2015}%
  \BibitemOpen
  \bibfield  {author} {\bibinfo {author} {\bibfnamefont {Y.-H.}\ \bibnamefont
  {Sham}}, \bibinfo {author} {\bibfnamefont {T.~K.}\ \bibnamefont {Chan}},
  \bibinfo {author} {\bibfnamefont {L.-M.}\ \bibnamefont {Lin}}, \ and\
  \bibinfo {author} {\bibfnamefont {P.~T.}\ \bibnamefont {Leung}},\ }\href
  {\doibase 10.1088/0004-637x/798/2/121} {\bibfield  {journal} {\bibinfo
  {journal} {The Astrophysical Journal}\ }\textbf {\bibinfo {volume} {798}},\
  \bibinfo {pages} {121} (\bibinfo {year} {2015})}\BibitemShut {NoStop}%
\bibitem [{\citenamefont {Chan}\ \emph {et~al.}(2015)\citenamefont {Chan},
  \citenamefont {Chan},\ and\ \citenamefont {Leung}}]{PhysRevD.91.044017}%
  \BibitemOpen
  \bibfield  {author} {\bibinfo {author} {\bibfnamefont {T.~K.}\ \bibnamefont
  {Chan}}, \bibinfo {author} {\bibfnamefont {A.~P.~O.}\ \bibnamefont {Chan}}, \
  and\ \bibinfo {author} {\bibfnamefont {P.~T.}\ \bibnamefont {Leung}},\ }\href
  {\doibase 10.1103/PhysRevD.91.044017} {\bibfield  {journal} {\bibinfo
  {journal} {Phys. Rev. D}\ }\textbf {\bibinfo {volume} {91}},\ \bibinfo
  {pages} {044017} (\bibinfo {year} {2015})}\BibitemShut {NoStop}%
\bibitem [{\citenamefont {Tolman}(1939)}]{PhysRev.55.364}%
  \BibitemOpen
  \bibfield  {author} {\bibinfo {author} {\bibfnamefont {R.~C.}\ \bibnamefont
  {Tolman}},\ }\href {\doibase 10.1103/PhysRev.55.364} {\bibfield  {journal}
  {\bibinfo  {journal} {Phys. Rev.}\ }\textbf {\bibinfo {volume} {55}},\
  \bibinfo {pages} {364} (\bibinfo {year} {1939})}\BibitemShut {NoStop}%
\bibitem [{\citenamefont {Jiang}\ and\ \citenamefont
  {Yagi}(2019)}]{PhysRevD.99.124029}%
  \BibitemOpen
  \bibfield  {author} {\bibinfo {author} {\bibfnamefont {N.}~\bibnamefont
  {Jiang}}\ and\ \bibinfo {author} {\bibfnamefont {K.}~\bibnamefont {Yagi}},\
  }\href {\doibase 10.1103/PhysRevD.99.124029} {\bibfield  {journal} {\bibinfo
  {journal} {Phys. Rev. D}\ }\textbf {\bibinfo {volume} {99}},\ \bibinfo
  {pages} {124029} (\bibinfo {year} {2019})}\BibitemShut {NoStop}%
\bibitem [{git()}]{github}%
  \BibitemOpen
  \href@noop {} {}\bibinfo {note}
  {{\url{https://github.com/nj2nu/I-C_Love-C_6thorderSeriesExpansion_Coefficients}}}\BibitemShut
  {NoStop}%
\bibitem [{\citenamefont {Maselli}\ \emph {et~al.}(2013)\citenamefont
  {Maselli}, \citenamefont {Cardoso}, \citenamefont {Ferrari}, \citenamefont
  {Gualtieri},\ and\ \citenamefont {Pani}}]{Maselli:2013mva}%
  \BibitemOpen
  \bibfield  {author} {\bibinfo {author} {\bibfnamefont {A.}~\bibnamefont
  {Maselli}}, \bibinfo {author} {\bibfnamefont {V.}~\bibnamefont {Cardoso}},
  \bibinfo {author} {\bibfnamefont {V.}~\bibnamefont {Ferrari}}, \bibinfo
  {author} {\bibfnamefont {L.}~\bibnamefont {Gualtieri}}, \ and\ \bibinfo
  {author} {\bibfnamefont {P.}~\bibnamefont {Pani}},\ }\href {\doibase
  10.1103/PhysRevD.88.023007} {\bibfield  {journal} {\bibinfo  {journal} {Phys.
  Rev. D}\ }\textbf {\bibinfo {volume} {88}},\ \bibinfo {pages} {023007}
  (\bibinfo {year} {2013})},\ \Eprint {http://arxiv.org/abs/1304.2052}
  {arXiv:1304.2052 [gr-qc]} \BibitemShut {NoStop}%
\bibitem [{\citenamefont {Urbanec}\ \emph {et~al.}(2013)\citenamefont
  {Urbanec}, \citenamefont {Miller},\ and\ \citenamefont
  {Stuchlik}}]{Urbanec:2013fs}%
  \BibitemOpen
  \bibfield  {author} {\bibinfo {author} {\bibfnamefont {M.}~\bibnamefont
  {Urbanec}}, \bibinfo {author} {\bibfnamefont {J.~C.}\ \bibnamefont {Miller}},
  \ and\ \bibinfo {author} {\bibfnamefont {Z.}~\bibnamefont {Stuchlik}},\
  }\href {\doibase 10.1093/mnras/stt858} {\bibfield  {journal} {\bibinfo
  {journal} {Mon. Not. Roy. Astron. Soc.}\ }\textbf {\bibinfo {volume} {433}},\
  \bibinfo {pages} {1903} (\bibinfo {year} {2013})},\ \Eprint
  {http://arxiv.org/abs/1301.5925} {arXiv:1301.5925 [astro-ph.SR]} \BibitemShut
  {NoStop}%
\bibitem [{\citenamefont {{Hartle}}(1967)}]{1967ApJ150.1005H}%
  \BibitemOpen
  \bibfield  {author} {\bibinfo {author} {\bibfnamefont {J.~B.}\ \bibnamefont
  {{Hartle}}},\ }\href {\doibase 10.1086/149400} {\bibfield  {journal}
  {\bibinfo  {journal} {\apj}\ }\textbf {\bibinfo {volume} {150}},\ \bibinfo
  {pages} {1005} (\bibinfo {year} {1967})}\BibitemShut {NoStop}%
\bibitem [{\citenamefont {Breu}\ and\ \citenamefont
  {Rezzolla}(2016)}]{Breu:2016ufb}%
  \BibitemOpen
  \bibfield  {author} {\bibinfo {author} {\bibfnamefont {C.}~\bibnamefont
  {Breu}}\ and\ \bibinfo {author} {\bibfnamefont {L.}~\bibnamefont
  {Rezzolla}},\ }\href {\doibase 10.1093/mnras/stw575} {\bibfield  {journal}
  {\bibinfo  {journal} {Mon. Not. Roy. Astron. Soc.}\ }\textbf {\bibinfo
  {volume} {459}},\ \bibinfo {pages} {646} (\bibinfo {year} {2016})},\ \Eprint
  {http://arxiv.org/abs/1601.06083} {arXiv:1601.06083 [gr-qc]} \BibitemShut
  {NoStop}%
\bibitem [{\citenamefont {Staykov}\ \emph {et~al.}(2016)\citenamefont
  {Staykov}, \citenamefont {Doneva},\ and\ \citenamefont
  {Yazadjiev}}]{Staykov:2016mbt}%
  \BibitemOpen
  \bibfield  {author} {\bibinfo {author} {\bibfnamefont {K.~V.}\ \bibnamefont
  {Staykov}}, \bibinfo {author} {\bibfnamefont {D.~D.}\ \bibnamefont {Doneva}},
  \ and\ \bibinfo {author} {\bibfnamefont {S.~S.}\ \bibnamefont {Yazadjiev}},\
  }\href {\doibase 10.1103/PhysRevD.93.084010} {\bibfield  {journal} {\bibinfo
  {journal} {Phys. Rev.}\ }\textbf {\bibinfo {volume} {D93}},\ \bibinfo {pages}
  {084010} (\bibinfo {year} {2016})},\ \Eprint
  {http://arxiv.org/abs/1602.00504} {arXiv:1602.00504 [gr-qc]} \BibitemShut
  {NoStop}%
\bibitem [{\citenamefont {Damour}\ and\ \citenamefont
  {Nagar}(2009)}]{Damour:2009vw}%
  \BibitemOpen
  \bibfield  {author} {\bibinfo {author} {\bibfnamefont {T.}~\bibnamefont
  {Damour}}\ and\ \bibinfo {author} {\bibfnamefont {A.}~\bibnamefont {Nagar}},\
  }\href {\doibase 10.1103/PhysRevD.80.084035} {\bibfield  {journal} {\bibinfo
  {journal} {Phys. Rev.}\ }\textbf {\bibinfo {volume} {D80}},\ \bibinfo {pages}
  {084035} (\bibinfo {year} {2009})},\ \Eprint {http://arxiv.org/abs/0906.0096}
  {arXiv:0906.0096 [gr-qc]} \BibitemShut {NoStop}%
\bibitem [{\citenamefont {{Boshkayev}}\ \emph {et~al.}(2017)\citenamefont
  {{Boshkayev}}, \citenamefont {{Quevedo}},\ and\ \citenamefont
  {{Zhami}}}]{ilq_wd}%
  \BibitemOpen
  \bibfield  {author} {\bibinfo {author} {\bibfnamefont {K.}~\bibnamefont
  {{Boshkayev}}}, \bibinfo {author} {\bibfnamefont {H.}~\bibnamefont
  {{Quevedo}}}, \ and\ \bibinfo {author} {\bibfnamefont {B.}~\bibnamefont
  {{Zhami}}},\ }\href {\doibase 10.1093/mnras/stw2614} {\bibfield  {journal}
  {\bibinfo  {journal} {Mon. Not. Roy. Astron. Soc.}\ }\textbf {\bibinfo
  {volume} {464}},\ \bibinfo {pages} {4349} (\bibinfo {year}
  {2017})}\BibitemShut {NoStop}%
\bibitem [{\citenamefont {{Boshkayev}}\ and\ \citenamefont
  {{Quevedo}}(2018)}]{hotwd}%
  \BibitemOpen
  \bibfield  {author} {\bibinfo {author} {\bibfnamefont {K.}~\bibnamefont
  {{Boshkayev}}}\ and\ \bibinfo {author} {\bibfnamefont {H.}~\bibnamefont
  {{Quevedo}}},\ }\href {\doibase 10.1093/mnras/sty1227} {\bibfield  {journal}
  {\bibinfo  {journal} {Mon. Not. Roy. Astron. Soc.}\ }\textbf {\bibinfo
  {volume} {478}},\ \bibinfo {pages} {1893} (\bibinfo {year} {2018})},\ \Eprint
  {http://arxiv.org/abs/1709.04593} {arXiv:1709.04593 [astro-ph.SR]}
  \BibitemShut {NoStop}%
\bibitem [{\citenamefont {Taylor}\ \emph {et~al.}(2020)\citenamefont {Taylor},
  \citenamefont {Yagi},\ and\ \citenamefont {Arras}}]{Taylor:2019hle}%
  \BibitemOpen
  \bibfield  {author} {\bibinfo {author} {\bibfnamefont {A.}~\bibnamefont
  {Taylor}}, \bibinfo {author} {\bibfnamefont {K.}~\bibnamefont {Yagi}}, \ and\
  \bibinfo {author} {\bibfnamefont {P.}~\bibnamefont {Arras}},\ }\href
  {\doibase 10.1093/mnras/stz3519} {\bibfield  {journal} {\bibinfo  {journal}
  {Mon. Not. Roy. Astron. Soc.}\ }\textbf {\bibinfo {volume} {492}},\ \bibinfo
  {pages} {978} (\bibinfo {year} {2020})},\ \Eprint
  {http://arxiv.org/abs/1912.09557} {arXiv:1912.09557 [gr-qc]} \BibitemShut
  {NoStop}%
\bibitem [{\citenamefont {Petroff}(2007)}]{Petroff:2007tz}%
  \BibitemOpen
  \bibfield  {author} {\bibinfo {author} {\bibfnamefont {D.}~\bibnamefont
  {Petroff}},\ }\href {\doibase 10.1088/0264-9381/24/5/003} {\bibfield
  {journal} {\bibinfo  {journal} {Class. Quant. Grav.}\ }\textbf {\bibinfo
  {volume} {24}},\ \bibinfo {pages} {1055} (\bibinfo {year} {2007})},\ \Eprint
  {http://arxiv.org/abs/gr-qc/0701081} {arXiv:gr-qc/0701081 [gr-qc]}
  \BibitemShut {NoStop}%
\bibitem [{\citenamefont {Stein}\ \emph {et~al.}(2014)\citenamefont {Stein},
  \citenamefont {Yagi},\ and\ \citenamefont {Yunes}}]{Stein:2013ofa}%
  \BibitemOpen
  \bibfield  {author} {\bibinfo {author} {\bibfnamefont {L.~C.}\ \bibnamefont
  {Stein}}, \bibinfo {author} {\bibfnamefont {K.}~\bibnamefont {Yagi}}, \ and\
  \bibinfo {author} {\bibfnamefont {N.}~\bibnamefont {Yunes}},\ }\href
  {\doibase 10.1088/0004-637X/788/1/15} {\bibfield  {journal} {\bibinfo
  {journal} {Astrophys. J.}\ }\textbf {\bibinfo {volume} {788}},\ \bibinfo
  {pages} {15} (\bibinfo {year} {2014})},\ \Eprint
  {http://arxiv.org/abs/1312.4532} {arXiv:1312.4532 [gr-qc]} \BibitemShut
  {NoStop}%
\bibitem [{\citenamefont {Yagi}\ \emph
  {et~al.}(2014{\natexlab{b}})\citenamefont {Yagi}, \citenamefont {Kyutoku},
  \citenamefont {Pappas}, \citenamefont {Yunes},\ and\ \citenamefont
  {Apostolatos}}]{Yagi:2014bxa}%
  \BibitemOpen
  \bibfield  {author} {\bibinfo {author} {\bibfnamefont {K.}~\bibnamefont
  {Yagi}}, \bibinfo {author} {\bibfnamefont {K.}~\bibnamefont {Kyutoku}},
  \bibinfo {author} {\bibfnamefont {G.}~\bibnamefont {Pappas}}, \bibinfo
  {author} {\bibfnamefont {N.}~\bibnamefont {Yunes}}, \ and\ \bibinfo {author}
  {\bibfnamefont {T.~A.}\ \bibnamefont {Apostolatos}},\ }\href {\doibase
  10.1103/PhysRevD.89.124013} {\bibfield  {journal} {\bibinfo  {journal} {Phys.
  Rev.}\ }\textbf {\bibinfo {volume} {D89}},\ \bibinfo {pages} {124013}
  (\bibinfo {year} {2014}{\natexlab{b}})},\ \Eprint
  {http://arxiv.org/abs/1403.6243} {arXiv:1403.6243 [gr-qc]} \BibitemShut
  {NoStop}%
\bibitem [{\citenamefont {Majumder}\ \emph {et~al.}(2015)\citenamefont
  {Majumder}, \citenamefont {Yagi},\ and\ \citenamefont
  {Yunes}}]{Majumder:2015kfa}%
  \BibitemOpen
  \bibfield  {author} {\bibinfo {author} {\bibfnamefont {B.}~\bibnamefont
  {Majumder}}, \bibinfo {author} {\bibfnamefont {K.}~\bibnamefont {Yagi}}, \
  and\ \bibinfo {author} {\bibfnamefont {N.}~\bibnamefont {Yunes}},\ }\href
  {\doibase 10.1103/PhysRevD.92.024020} {\bibfield  {journal} {\bibinfo
  {journal} {Phys. Rev.}\ }\textbf {\bibinfo {volume} {D92}},\ \bibinfo {pages}
  {024020} (\bibinfo {year} {2015})},\ \Eprint
  {http://arxiv.org/abs/1504.02506} {arXiv:1504.02506 [gr-qc]} \BibitemShut
  {NoStop}%
\bibitem [{\citenamefont {Yagi}(2014)}]{Yagi:2013sva}%
  \BibitemOpen
  \bibfield  {author} {\bibinfo {author} {\bibfnamefont {K.}~\bibnamefont
  {Yagi}},\ }\href {\doibase 10.1103/PhysRevD.97.129901,
  10.1103/PhysRevD.96.129904, 10.1103/PhysRevD.89.043011} {\bibfield  {journal}
  {\bibinfo  {journal} {Phys. Rev.}\ }\textbf {\bibinfo {volume} {D89}},\
  \bibinfo {pages} {043011} (\bibinfo {year} {2014})},\ \bibinfo {note}
  {[Erratum: Phys. Rev.D96,no.12,129904(2017); Erratum: Phys.
  Rev.D97,no.12,129901(2018)]},\ \Eprint {http://arxiv.org/abs/1311.0872}
  {arXiv:1311.0872 [gr-qc]} \BibitemShut {NoStop}%
\end{thebibliography}%
\end{document}